\documentclass[fleqn,usenatbib]{mnras}

\usepackage{newtxtext,newtxmath}

\usepackage[T1]{fontenc}

\DeclareRobustCommand{\VAN}[3]{#2}
\let\VANthebibliography\thebibliography
\def\thebibliography{\DeclareRobustCommand{\VAN}[3]{##3}\VANthebibliography}

\usepackage{graphicx}	
\usepackage{amsmath}	
\usepackage{url}
\usepackage{subcaption}

\defcitealias{lane22}{LBM22}

\newcommand{\Msun}{$\mathrm{M}_{\odot}$}

\title[DFs of simulated accretion remnants]{Distribution functions for the modelling of accretion remnants in Milky Way-like galaxies: insights from IllustrisTNG}

\author[Lane \& Bovy]{
James M. M. Lane$^{1}$\thanks{E-mail: jm.lane@alumni.utoronto.ca} \&
Jo Bovy$^{1}$\thanks{E-mail:
bovy@astro.utoronto.ca}
\\
$^{1}$David A. Dunlap Department of Astronomy and Astrophysics, University of Toronto, 50 St. George Street, Toronto ON, M5S 3H4, Canada\\
}

\date{Accepted XXX. Received YYY; in original form ZZZ}

\pubyear{2025}

\begin{document}
\label{firstpage}
\pagerange{\pageref{firstpage}--\pageref{lastpage}}
\maketitle

\begin{abstract}

We study accretion remnants around Milky Way analogs in the IllustrisTNG simulations to determine how well commonly used distribution functions (DFs) describe their phase-space distributions. We identify 30 Milky Way analogs and 116 remnants from mergers with stellar mass ratios greater than 1:20. Two-power density profiles, as well as rotating constant-anisotropy and Osipkov-Merritt DFs are fit to the remnants. We determine that the remnants are suitable for equilibrium modelling by assessing them in the context of the Jeans equation. Each of the models we consider are reasonably able to fit the stellar remnant energy and angular momentum distribution, as well as the magnitude and shape of velocity dispersion profiles. Case studies matched to two well-known merger remnants in the stellar halo---\textit{Gaia}-Sausage/Enceladus (GS/E) and Sequoia---are explored in more depth. We find good evidence that remnants with high anisotropy $\beta$, such as GS/E, are better modelled with a superposition of two Osipkov-Merritt DFs than either a constant-anisotropy model or a single Osipkov-Merritt DF. We estimate an Osipkov-Merritt profile with scale radius between $2-4$~kpc would be a good first-order representation of GS/E, and comment on existing observational evidence for this as well as studies which could demonstrate it. Overall, we find that DF-based models work well for describing the kinematics of large merger remnants. Our results will be an important reference for future studies which seek to constrain both the spatial and kinematic properties of merger remnants in the Milky Way stellar halo.

\end{abstract}




\section{Introduction}
\label{sec:introduction}

The stellar content of the Milky Way halo bears testament to a long history of accretion of smaller structures such as dwarf galaxies \citep{helmi20,deason24}. These accretion events are one of the principal growth modes for a typical spiral galaxy like the Milky Way in a $\Lambda$CDM context, and occur as a consequence of the fact that mass in our Universe is arranged hierarchically such that each galaxy is surrounded by large numbers of smaller structures \citep{white78,searle78,bullock05,cooper10}. Not only are mergers an important ingredient in the formation of a galaxy, but at later times they provide a substantial reservoir of matter, both baryonic and dark, for a growing galaxy and the dynamical impact can profoundly alter the kinematics and morphology of the host as well. 

While the fact that the Milky Way hosts numerous dwarf galaxy satellites has been long known, the observation of the tidal tails of the actively-merging Sagittarius dwarf galaxy provided a glimpse into the process of accretion at the present \citep{ibata94,belokurov06}. Ongoing observations reveal countless coherent stellar systems in various states of dissolution, from dwarf galaxies with barely visible tidal tails, to fully dissolved stellar streams, to the Magellanic clouds which are actively falling into the Milky Way. Together, these observations paint a vivid picture of the past, present, and future growth of the Galaxy.

The \textit{Gaia Space Telescope} \citep{gaia} has revolutionized our understanding of the stellar halo by providing accurate 5D phase space information for nearly 2 billion---and 6D phase space information for over 20 million---stars. Combining this astrometric data with abundances and stellar parameters from large ground-based spectroscopic campaigns has generated a comprehensive dataset to search for the remnants of ancient mergers in the Milky Way stellar halo. The discovery of a large population of metal-rich halo stars on highly eccentric orbits \citep{belokurov18,haywood18,helmi18}, now dubbed the \textit{Gaia}-Sausage/Enceladus (GS/E), which dominates the nearby stellar halo represents the first major finding of a phase-mixed merger remnant in this new era of data.

Besides GS/E a vast number of other diffuse stellar halo structures have been discovered. Many of these are supposed accretion remnants, observed as distinct groups of stars or globular clusters in chemodynamical space, such as the Helmi Streams \citep{helmi99,koppelman19a}, Thamnos \citep{koppelman19b}, Sequoia \citep{myeong19}, Heracles \citep{horta21}, as well as the numerous structures (Aleph, Arjuna, I'itoi, Wukong) reported by \citet{naidu20}. The Kraken \citep{kruijssen20} and the Koala \citep{forbes20} are inferred using the chemodynamics and ages of Milky Way globular clusters. Some \textit{in-situ} halo components related to the thick disk have also been identified, most notably the Splash \citep{belokurov20}. More recently, the old \textit{in-situ} stellar halo (dubbed Aurora) has been studied directly \citep{belokurov22,conroy22,rix22}, and its links to the `spin-up' of the Milky Way disk established.

With this extensive trove of structures identified in recent years, driven by new astrometric and spectroscopic data, the task now turns to characterizing them by modelling their properties. This typically consists of a detailed study of their chemistry and past chemical evolution \citep[e.g., for GS/E see][]{vincenzo19,monty20,hasselquist21}, which may motivate an estimate of the stellar mass of the progenitor. The modelling of kinematics by fitting density profiles or distribution functions is less common, but has been performed with some success for GS/E \citep{lancaster19,mackereth20,iorio21,han22,lane23}. Another approach is to use N-body techniques to either identify similar remnants in the halos of Milky Way analogs in a cosmological setting, or use tailored N-body models that seek to reproduce specific features of an observed remnant \citep[e.g.,][again for GS/E]{fattahi19,mackereth19a,naidu21,amarante22,orkney23}. Each of these approaches is complementary, and may be suitably applied to some halo structures but not others. It is regardless important to model the properties of these structures, to discover whether they are genuine accretion remnants or \textit{in-situ} components of the Milky Way, and if so to characterize their chemistry, kinematics, and spatial distribution. The converse is important as well, to discover whether a structure is not unique but perhaps some complex dynamical echo of another remnant \citep[see][]{jean-baptiste17}, or an artifact of selection effects \citep[see][]{lane22}.

In this work we turn our attention to distribution functions (DFs), perhaps the least-used technique of the aforementioned for modelling individual remnants. Density profiles have been fit to the whole stellar halo for some time \citep[for contemporary examples see][]{deason19,mackereth20}, and have recently been fit specifically to GS/E \citep{han22,lane23}. Multi-component Gaussian models, representing simple DFs, have been applied to kinematic data in order to study GS/E as well with good success \citep[e.g.][]{lancaster19,fattahi19,iorio21}. More recently, \citet{lane22} and \citet{lane23} employed simple constant-anisotropy DFs to study and fit density profiles to GS/E, however they did not actively fit the DF to data but constructed a DF based on assumptions about GS/E to aid in sample selection. It has become clear in the \textit{Gaia} era that remnants in the stellar halo have unique spatial, kinematic, and chemical properties, making DFs better suited than just density profiles owing to their ability to fold in kinematic information implicitly as well as chemistry with simple additions \citep[e.g.][]{sanders15b}. Furthermore, there are many DFs, including those that we propose to study in this work, which are better physically motivated than simpler models such as Gaussians, which are often preferred due to their flexibility and ubiquity. DFs have now been used to study many other dynamical groups in the Milky Way, including the stellar disk, bar, bulge, as well as the stellar halo as a whole, in addition to the dark halo. We believe that they can now also be applied to better study, classify, and understand the individual merger remnants that comprise the stellar halo as well. In particular, active fitting of DFs to data represents a next step in the study of major accretion remnants such as GS/E, but also smaller remnants and \textit{in-situ} structures.

In the context of simulations, the kinematics of accretion remnants has been studied in a broad sense \citep[e.g.][]{johnston08,deason13,amorisco17,jean-baptiste17}. But such studies do not typically focus directly on DFs, rather seek to gain an empirical understanding of kinematic properties. Additionally, the specific case of GS/E has been investigated with tailored simulations over the last few years owing to its prominence as a puzzle of the modern stellar halo \citep{naidu21,amarante22}. Recently, the stellar halos of simulated M31 analogs were studied by \citet{gherghinescu24} using action-based DFs, but they did not focus on individual remnants. We propose that a useful addition to this body of research would be to directly assess simulated accretion remnants using DFs to inform future efforts to model the Milky Way stellar halo on a remnant-by-remnant basis.

In this paper we will identify and study accretion remnants in the IllustrisTNG cosmological simulation \citep{tng_public_release_nelson19}, and investigate how well commonly used classes of DFs are able to fit the data. We focus specifically on a small number of well known DFs in this work aware that there are other forms and novel techniques that may be studied in the future, and will provide some discussion of these avenues at the end of the paper. Our overall aim for this work is to extract insight using these simple DFs to inform future work on the topic, but more importantly to provide a basis from which upcoming studies of remnants using DFs can start to proceed more confidently. We do not aim to provide definitive techniques for optimal DF fits to simulated data, but hope that this study may lay the ground work to accomplish this goal down the line. The paper is laid out as follows: \S~\ref{sec:simulations} provides an overview of the IllustrisTNG simulations and describes the selection of Milky Way analogs as well and major mergers. \S~\ref{sec:fitting-distribution-functions} describes the DFs that we use and provides the fitting methodology. In \S~\ref{sec:assessment-comparison-distribution-functions} we assess how well each DF model is matched to the data. We end with a discussion and conclusion in sections \S~\ref{sec:discussion} and \S~\ref{sec:summary-conclusions} respectively.

\section{Simulations}
\label{sec:simulations}

We study the high resolution TNG50 simulation \citep{tng50_nelson19,tng50_pillepich19} of the IllustrisTNG project \citep{tng_public_release_nelson19}. We specifically use the highest resolution TNG50-1 simulation which is contained in a box of 51.7 comoving Mpc$^{3}$ sampled by $2160^{3}$ each of gas cells, dark matter particles, and Monte Carlo tracers. The initial conditions of the simulations are cosmologically motivated, and based on the \citet{planck15} cosmology with a hubble parameter $H_{0} = 100\,h\,\mathrm{km\,s}^{-1}\,\mathrm{Mpc}^{-1} = 67.74 \,\mathrm{km\,s}^{-1}\,\mathrm{Mpc}^{-1} $. The simulations are evolved using the moving mesh code \texttt{AREPO} \citep{arepo_springel10}, with galaxy formation and physics models outlined by \citet{tng_model_weinberger17} and \citet{tng_model_pillepich18}. These models are calibrated to reproduce the $z=0$ galaxy stellar mass function, galaxy stellar-to-halo mass relation, and galaxy stellar mass-size relation, among other observed trends and relations less pertinant to this work.

The typical particle masses are $8.5 \times 10^{4}$~\Msun\ for baryons and $4.5\times10^{5}$~\Msun\ for dark matter. These small particle masses are crucial to be able to properly resolve the low-mass merger remnants, of order $10^{7}-10^{9}$~\Msun\ stellar mass, which are of greatest interest in the context of the Milky Way right now. The Plummer-equivalent gravitational softening length, $\epsilon_{\star}$, for collisionless interactions is equal to 0.39\,$h$~comoving~kpc (equal to 0.29~kpc at $z=0$ given $h$ defined above). Dark matter halos are identified using the friends-of-friends algorithm with linking length 0.2, and individual galaxies (subhalos) are identified using the \texttt{SUBFIND} algorithm \citep{subfind_springel01}.

\subsection{Selection of Milky Way analogs}
\label{subsec:analog-selection}

\begin{figure*}
    \centering
    \includegraphics[width=\textwidth]{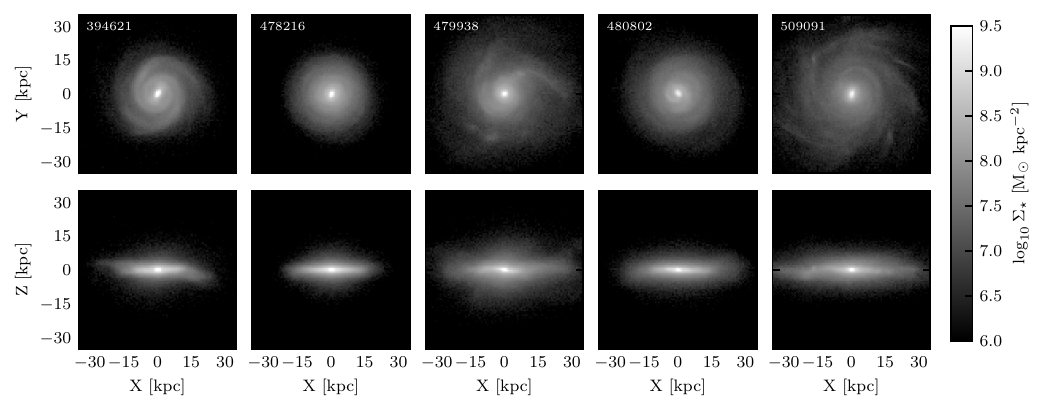}
    \caption{Stellar mass surface density for five of the Milky Way analogs selected in \S~\ref{subsec:analog-selection} shown face-on (top) and edge-on (bottom) at $z=0$. The subfind IDs of each galaxy is annotated in the top row of panels, and distances are in physical kpc.}
    \label{fig:galaxy-surface-density}
\end{figure*}

We select Milky Way analogs in the TNG50-1 simulation based on the $z=0$ stellar mass of subhalos identified using \texttt{SUBFIND}. We choose a stellar mass range of $5\times10^{10}$~\Msun$< M_{\star} < 7 \times 10^{10}$~\Msun, which brackets modern stellar mass estimates of the Milky Way \citep[e.g. see][]{bovy13,bland-hawthorn16}. We identify 46 galaxies in TNG50-1 within this stellar mass range, and a sample of five of them is shown in Figure~\ref{fig:galaxy-surface-density}.

Before continuing on with sample selection, we describe here some steps we take to prepare subhalos for kinematic analysis. First, we determine the center-of-mass of each subhalo by applying the recursive shrinking-spherical-center algorithm described by \citet{power03} to the star particles, shrinking the sphere by a factor 0.9 (i.e. each new radius is 0.9 times the size of the previous) each iteration and halting when fewer than 100 particles are enclosed. From particle velocities we subtract the mass-weighted mean velocity of all star particles within one stellar half-mass radius, $r_{h}$, of the center of the subhalo. We then rectify the galaxy to the vector described by the total angular momentum of the star particles between $0.5-2$ times $r_{h}$ from the center.

Now returning to analog selection, because our simple selection for Milky Way analogs by their stellar mass is agnostic towards the morphology of the chosen galaxies, we further pare down the sample by examining the relative fraction of spherical (bulge and halo) and disky stellar components. We assess these quantities using the orbit circularity, defined as $\eta = J_{z}/J_\mathrm{circ}(E)$ the specific angular momentum about the z-axis divided by the specific angular momentum of a circular orbit with a given energy $E$. We determine $J_\mathrm{circ}(E)$ empirically by relying on the fact that at any energy the maximum allowed angular momentum is that of a circular orbit. We therefore calculate the maximum angular momentum for particles in 100 bins evenly spanning the minimum to maximum energies, and fitting to this trend a 1-D spline, which gives $J_\mathrm{circ}(E)$.

\begin{figure}
    \centering
    \includegraphics[width=\columnwidth]{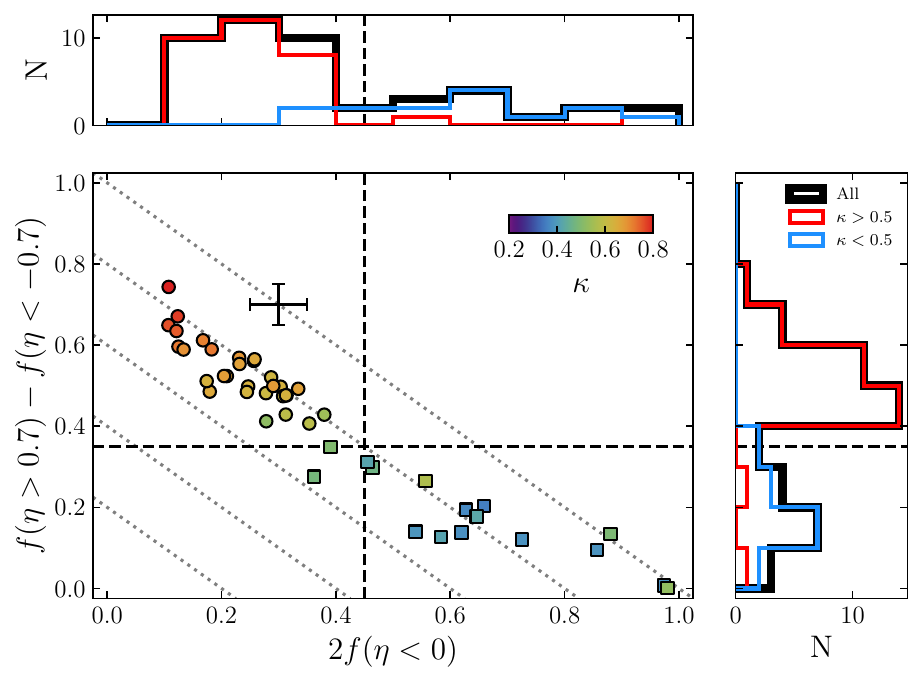}
    \caption{The disk proxy (fraction of circularities above 0.7 minus the fraction below $-0.7$) as a function of the spheroid proxy (twice the fraction of circularities below 0) for the 46 mass-selected analogs. Histograms in the right and top panels show the margins of each of these quantities respectively. Circles are analogs included in the sample based on cuts, and squares are analogs excluded from the sample. The colorscale shows the fraction of energy in ordered rotation, $\kappa$. The dashed lines show the cuts adopted to separate galaxies with Milky Way-like spheroid and disk fractions. Those remaining in the sample are circles, and those removed from the sample are crosses. The set of error bars in the top-left quadrant shows the approximate bulge and disk fraction of the Milky Way.}
    \label{fig:bulge-disk-decomposition}
\end{figure}

To proxy the disk we consider the fraction of stars within $2r_{h}$ with circularity greater than 0.7, subtracting the fraction of stars within $2r_{h}$ with circularity less than $-0.7$ (hereafter disk fraction). For the bulge or spherical component we use twice the fraction of stars within $2r_{h}$ with circularity less than 0 (hereafter bulge fraction). Figure~\ref{fig:bulge-disk-decomposition} shows these quantities for the 46 stellar mass-selected analogs. The marginal distribution of these quantities reveals two populations, a concentrated group with spheroid and disk fractions between $0.1-0.4$ and $0.4-0.8$ respectively, alongside a second group or extended tail of the distribution at larger spheroid fraction and lower disk fraction. Visual examination reveals that many of the galaxies in the group with higher spheroid fraction are undergoing mergers at the present day.

The population with higher disk fraction is broadly consistent with the Milky Way, which has an approximate stellar bulge mass fraction of 0.3 and disk mass fraction of about 0.7 \citep[see][and references therein]{bland-hawthorn16}. This set of mass fractions, with fiducial uncertainties of 0.05, is shown on Figure~\ref{fig:bulge-disk-decomposition}. Note that the population of Milky Way analogs does not overlap this locus, and we do not expect it to because the Milky Way mass fraction values are determined in a much more systematic, tailored manner, whereas our metrics are more approximate.

We extract the 30 galaxies with disk fraction greater than 0.35 and spheroid fraction less than 0.45, and remove the 16 other galaxies from the sample. There is good overlap between our sample and the sample of TNG50 Milky Way analogs compiled by \citet{tng50_mw_pillepich24}. All but one of our analogs is also found in their much larger (numbering 198) sample of analogs which is defined by a much broader stellar mass range ($10^{10.5} \sim 3\times10^{10} < M_{\star} / {\rm M}_{\odot} < 10^{11.2} \sim 1.5\times10^{11}$), as well as an isolation criterion and a less stringent morphology criterion based on the aspect ratios of the stellar light distribution.

We compare this selection using proxied disk and spheroid fractions to other types of selections which can discriminate between galaxies with disky and spheroidal stellar components. First we consider the fraction of energy in ordered rotation \citep{sales10,correa17}, defined as 
\begin{equation}
    \label{eq:kappa}
    \kappa = \frac{ \sum_{i}^{N} m_{i} (J_{z,i}/R_{i})^{2}  }{ \sum_{i}^{N} m_{i} v_{i}^{2} }\,,
\end{equation}
\noindent where $J_{z,i}$ are specific angular momenta about the disk axis, $R_{i}$ are cylindrical radii, $v_{i}$ is the velocity magnitude for each particle, and $m_{i}$ are particle masses. We compute $\kappa$ for all star particles within $2r_{h}$, and report the values as the colorscale in Figure~\ref{fig:bulge-disk-decomposition}. It is clear when examining this figure that $\kappa$ trends strongly with more Milky Way-like disk and spheroid fraction. Indeed when examining $\kappa$ alone we find two groups at approximately $\kappa=0.4$ and $\kappa=0.7$, and the sample very neatly divides at $\kappa=0.5$, with only two excluded analogs lying at slightly higher values, and all included analogs lying at higher values.

We also examine the bulge and disk fractions reported by \citet{du20} based on the approach presented by \citet{du19}. These authors perform an unsupervised decomposition of IllustrisTNG galaxies into traditional stellar components in the space of orbital circularity $\eta$ and normalized energy $E/\max(\vert E \vert)$. We find the galaxies remaining in our sample tend to have disk fraction $>0.4$ and bulge fraction between 0.05 and 0.4, while the galaxies removed from the sample have slightly higher bulge fractions between 0.1 and 0.5, but much lower disk fractions $<0.5$. Each of these two subsequent lines of analysis on our sample strengthen our confidence in the way in which we select our final analog sample. 

\subsection{Identification of major mergers}

For each Milky Way analog we identify major mergers of interest using merger trees generated with the \texttt{SUBLINK} \citep{rodriguez-gomez15} algorithm. In this work we are interested only in remnants which represent large accretion events for two reasons: first these remnants will be similar to the most thoroughly studied of the Milky Way remnants, and second these will have sufficient particle numbers such that kinematic modelling is possible. One has multiple options to define merger significance, it could be by present-day stellar or dark matter mass, or it could be by mass ratio at the epoch of the merger event. We elect to use the latter, because it allows for study of merger remnants from a variety of epochs on a more even footing.

\begin{figure}
    \centering
    \includegraphics{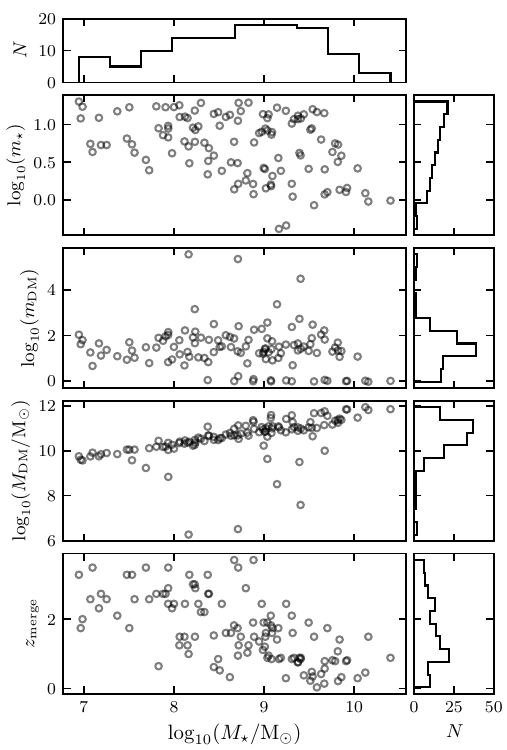}
    \caption{The physical properties and merger attributes of identified major mergers. The four main panels show, in descending order, primary-to-secondary stellar ($m_{\star}$) and dark matter ($m_{\rm DM}$) mass ratio, secondary dark matter mass, and merger redshift, all as functions of the secondary stellar mass. The mass ratios are computed at the time of maximum secondary stellar mass. The merger redshift is defined by the snapshot where the secondary is last recorded as an independent subhalo. The top-most panel shows the distribution of secondary stellar masses, and the four right-most panels show the distributions of each of the dependent parameters shown in the main panels.}
    \label{fig:merger-params}
\end{figure}

We broadly follow the approach of \citet{rodriguez-gomez16} when selecting merger remnants based on mass ratio. First, we consider the stellar mass ratio of the secondary versus the primary (the Milky Way analog). We choose the stellar mass ratio as opposed to the dark matter mass ratio because the star particles are more strongly bound to both the secondary and primary in the time leading up to the coalescence of the secondary, and so suffer less from pathological swapping and misplacing of particles by the \texttt{SUBFIND} algorithm during these close encounters. We also choose to measure the mass ratio at the time when the mass of the secondary reaches its largest value. We record all mergers with mass ratio greater than 1:20 at the corresponding time of largest secondary stellar mass. In this way we select 116 major mergers. An overview of their stellar and dark matter masses and primary-to-secondary mass ratios, as well as the redshift of the merger is shown in Figure~\ref{fig:merger-params}.

\section{Fitting distribution function models to data}
\label{sec:fitting-distribution-functions}

Here we first describe the three DF models that we use, and then the methods used to fit them to N-body remnant data. Throughout this section and the next we will on multiple occasions bin the data to compute velocity dispersions as well as the orbital anisotropy. In general we bin star particles according to the following scheme, unless specified otherwise. Bins for each remnant are of variable width and always contain $\min[500,N/10]$ particles, where $N$ is the total number of particles. In other words we use 500 particles per bin unless there are fewer than 5000 particles in the remnant ($\sim 4\times10^{8}$~\Msun\ remnant stellar mass), in which case we use 10 bins. The innermost bin edge always lies at $\max[\epsilon_{\star},\min(\{ r_{i} \})]$ where $\{r_{i}\}$ is the set of star particle galactocentric radii. All bins contain the same number of star particles, and particles outside the final bin edge are ignored. Our goal in using a fixed-number binning approach is to ensure that dispersions and anisotropies have consistent uncertainties, which depend on the number of star particles used for the computation.

\subsection{The constant-anisotropy and Osipkov-Merritt distribution functions}
\label{subsec:distribution-function-models}

The first model we consider is spherical with constant anisotropy, and takes the form
\begin{equation}
    \label{eq:constant-anisotropy-df}
    f(\mathcal{E},L) = L^{-2\beta} f_{1}(\mathcal{E})\,.
\end{equation}
\noindent Here the relative energy $\mathcal{E} = \Psi - \frac{1}{2}v^{2}$ is a function of $\Psi = -\Phi-\Phi(\infty)$, the negative of the gravitational potential offset to be 0 at infinity, as well as the velocity $v$. $L$ is the total angular momentum, and the parameter $\beta$ is the anisotropy, which is given as
\begin{equation}
    \label{eq:beta}
    \beta = 1- \frac{\sigma^{2}_{\theta} + \sigma^{2}_{\phi}}{2\sigma^{2}_{r}}\,,
\end{equation}
\noindent where $\sigma_{[\theta,\phi,r]}$ are the spherical velocity dispersions in the polar, azimuthal, and radial directions respectively. $\beta$ has the domain $(-\infty,1]$, with negative values corresponding to a tangentially biased system, a value of 0 corresponding to an isotropic or ergodic (i.e. the DF depends only on energy) system, and positive values corresponding to a radially biased system. Finally, the function $f_{1}$ relates the tracer density of the DF to the underlying potential. For ergodic ($\beta=0$) DFs $f_{1}$ can be solved by a simple integral inversion \citep{eddington16}, however for non-zero $\beta$ the calculation is more involved. For more information see \citet{cuddeford91}, \citet{an06a}, \citet{evans06}, chapter 4.3 of \citet{binney08}, and Appendix A of \citet{lane22} which specifically describes the implementation of this DF we use in \texttt{galpy} \citep{bovy15}. 

The second model we consider is typically independently attributed to \citet{osipkov79} and \cite{merritt85}, and is spherical with an anisotropy which approaches $\beta = 0$ at small radii, increasing with radii until it approaches $\beta = 1$ at large radii, reflecting the behaviour of many real galaxies. This DF takes the simple form
\begin{equation}
    \label{eq:osipkov-merritt-df}
    f(\mathcal{E},L) = f(Q)\,,
\end{equation}
\noindent where
\begin{equation}
    \label{eq:osipkov-merritt-Q}
    Q = \mathcal{E} - \frac{L^{2}}{2r_{a}^{2}},
\end{equation}
\noindent is a pseudo-energy which is constructed from the relative energy $\mathcal{E}$ and the total angular momentum $L$ scaled by a characteristic radius $r_{a}$. $f(Q)$ is computed via an integral inversion, in a manner similar to the ergodic and constant-anisotropy distribution functions. The anisotropy of the Osipkov-Merritt model varies as a function of spherical radius $r$, taking the form
\begin{equation}
    \label{eq:beta-osipkov-merritt}
    \beta(r) = \frac{r^{2}}{r^{2} + r_{a}^{2}}\,.
\end{equation}
\noindent From which it is clear that $\beta(0)=0$, that $\beta(r_{a})=1/2$, and that $\beta$ tends towards 1 as $r \rightarrow \infty$.

A final DF that we consider is a linear superposition of standard Osipkov-Merritt DFs with differing $r_{a}$ values. These types of DFs were first described by \citet{merritt85}, who noted that they allow for a wider range of anisotropy and velocity dispersion profiles than the standard single-$r_{a}$ form. A two-component Osipkov-Merritt superposition DF is defined as
\begin{equation}
    \label{eq:osipkov-merritt-superposition-df}
    f(Q) = k_\mathrm{om}f_{1}(Q) + (1-k_\mathrm{om})f_{2}(Q),
\end{equation}
\noindent where $f_{1}$ and $f_{2}$ are two standard Osipkov-Merritt DFs described by $r_{a,1}$ and $r_{a,2}$ and $k_\mathrm{om}$ is the fraction of the DF corresponding to $f_{1}$. Note that it is permissible for $f_{1}$ and $f_{2}$ to be defined by different tracer density profiles, which combine to produce the density profile of the remnant. Here we assume for simplicity that the density profiles of $f_{1}$ and $f_{2}$ are the same and equal to the best-fitting remnant density profile. It is also possible to explore a broader family of superposition DFs with additional linear components, or even with functional definitions for the weights corresponding to different $r_{a}$ and density profiles, but for simplicity we consider only models with two components here.

\subsection{The rotating DF prescription}
\label{subsec:rotating-df-prescription}

A final ingredient we add to our DFs is rotation about the galactic z-axis, which is implemented following the prescription of \citet{binney14d} \citep[see also][for an equivalent implementation using constant-anisotropy DFs]{deason11}. Consider that a DF $f(\mathbf{I})$, which is a function of some conserved quantities $\mathbf{I}$ such as energy and angular momentum or actions, can be split into an even (non-rotating) and odd (rotating) DF as 
\begin{equation}
    \label{eq:even-odd-df}
    f(\mathbf{I}) = (1-k_\mathrm{rot})f_{+}(\mathbf{I}) + k_\mathrm{rot}f_{-}(\mathbf{I})\,,
\end{equation}
\noindent where $k_\mathrm{rot}$ is the fraction of the DF which is rotating. If $f_{+}(\mathbf{I})$ is one of the DF models listed above, the odd portion of the DF is defined as
\begin{equation}
    \label{eq:rotating-df-kernel}
    f_{-}(\mathbf{I}) = \tanh \big( L_{\mathrm{z}}/\chi \big) f_{+}(\mathbf{I})\,,
\end{equation}
\noindent where $L_{\mathrm{z}}$ is the angular momentum about the galactic z-axis, and $\chi$ is the angular momentum scale at which rotation becomes significant. Because this prescription for rotation does not change the overall angular momentum $L$, or the energy $\mathcal{E}$ it can be applied to either the constant-anisotropy DF or the Osipkov-Merritt without changing the fundamental nature of the DF.

\subsection{Fitting distribution functions to data}
\label{subsec:fitting-constant-anisotropy-osipkov-merritt-dfs}

The procedure to fit the constant anisotropy and both standard and superposition Osipkov-Merritt DFs is similar. We consider that the required ingredients to construct these DFs are the gravitational potential $\Phi$, the density profile of the tracer $\rho$, and a parameter which specifies the anisotropy: $\beta$ for the constant-anisotropy DF, $r_{a}$ for the Osipkov-Merritt DF, and $\{ r_{a,1},r_{a,2},k_\mathrm{om} \}$ for the superposition Osipkov-Merritt DF. The choice to fit DFs in this way, using the density profile, underlying potential, and anisotropy-specifying parameters, is not the only way to accomplish this task. Since we have energy and angular momenta for the remnant particles we could fit the DF directly in the space of $(\mathcal{E},L)$ using either a custom form or the constant anisotropy and Osipkov-Merritt DFs. While this is possible, by focusing on the density profile and anisotropy profile---as we do---we are grounding the computed DFs in those observables, which we argue are just as important if not more important than the specific distribution in ($\mathcal{E},L$). DFs computed in this way are guaranteed to reproduce (at least to be very close, see below) the target density profile and anisotropy parameters. Additionally, while we do not attempt to prove this, we suspect that to fit the DF piece-wise as we do would produce very similar results to a DF fitted to integrals of motion. Finally it is relevent that if we were to fit the DF in $(\mathcal{E},L)$ space using the constant anisotropy or Osipkov-Merritt forms then it would add significant computational expense to the fitting process, as computation of such DFs is time consuming.

Note that, as derived by \citet{an06b}, there is a limit on the value of $\beta$ in the inner part of an anisotropic DF such that $\beta$ must be less than half the inner power law slope of the target density profile. This limit is applicable for self-consistent DFs, and is modified by a small factor when the potential and density and independent as is the case for the DFs used here. We recognize this limitation and navigate it in practice by setting any negative regions of computed DFs to 0 \citep[see][for more information on how this is done during construction of the DF]{lane22}. Such a choice means that any constructed DF will have density and anisotropy profile slightly different than are specified by the input ingredients, but the compromise ensures the physicality of the DF. Additionally, we only ever work with DFs at radii larger than the softening length (0.29~kpc) which slightly mitigates the impact of this effect.

Finally, here we neglect triaxiality in both the remnants and the potential. The host potentials are certainly expected to be flattened, and potentially triaxial in a cosmological context. In the inner parts of the system the baryonic disk flattens the potential, and the dark matter halo, which dominates the outskirts, is also expected to be flattened and mildly triaxial \citep{dubinski94,abadi10}. In order to assess the triaxiality of our sample we compute the moment of inertia tensor for each remmant. Diagonalizing the tensor gives the eigenvalues, which are the squares of the axis scale lengths $(a > b > c)$. We assess the distributions of $b/a$ and $c/a$, finding them peaked near their respective medians of 0.91 and 0.84. The distribution of $b/a$ ranges between 0.8 and 1, while $c/a$ ranges between 0.7 and 0.95.

So we see that most remnants are triaxial to a small or modest degree, indicating that a spherical assumption in our analysis is reasonable. Additionally, proper treatment of triaxiality is challenging. This is primarily because of the lack of good, general DFs with independent density profiles and potentials, but also because obtaining reliable integrals of motion for such systems is challenging. So here we work with stricly spherical models, and leave triaxiality to follow up studies. Additional discussion of the impact of this choice will be presented in \S~\ref{subsubsec:consideration-for-other-dfs}.

We begin by considering the host gravitational potential $\Phi$. Because each of the DF models relies on an explicitly spherical underlying potential (they are functions of $\mathcal{E}$ and total $L$), we construct a sphericalized representation of the mass profile of each host galaxy at $z=0$ to act as the potential. We first determine the enclosed mass profile $M(r)$ as a function of radius for all baryonic and dark matter particles. We then spline-interpolate the radial force $F_{r} = -\frac{G M(r)}{r^{2}}$ on a logarithmically-spaced grid from the minimum to maximum radius of all particles considered. From this representation the potential is computed trivially as the integral of the spline.

The second step is to fit the density of the tracer population, in this case each individual merger remnant. For the density profile we elect to use a two-power model which takes the form
\begin{equation}
    \label{eq:two-power-spherical-density}
    \rho_{\rm rem}(r) = \frac{ \rho_{\mathrm{r},0} }{ (r/r_{s})^{\alpha_{1}} (1+r/r_{s})^{\alpha_{2}-\alpha_{1}} }\,,
\end{equation}
\noindent where $\rho_{0}$ is the amplitude, $\alpha_{[1,2]}$ are the inner and outer power law indices, and $r_{s}$ is the scale radius marking the transition between the two power law indices. We choose this profile due to its flexibility, specifically in being able to realistically describe remnants such as GS/E, which is known to have a shallow inner density profile which steepens at a characteristic radius \citep[e.g.][]{han22,lane23}.

We fit this density profile to data using a likelihood-based framework, which we describe here very generally. At the core is Bayes theorem
\begin{equation}
    \label{eq:bayes-theorem}
    p(\mathbf{\Theta} \vert \{ \mathbf{X} \} ) \propto p( \{ \mathbf{X} \} \vert \mathbf{\Theta} ) p( \mathbf{\Theta} )
\end{equation}
\noindent where $\mathbf{\Theta}$ are a vector of model parameters to be determined and $\{ \mathbf{X} \}$ is a vector of data in a relevant coordinate frame. The posterior, $p(\mathbf{\Theta} \vert \{ \mathbf{X} \} )$, is of interest and is determined by the likelihood $p( \{ \mathbf{X} \} \vert \mathbf{\Theta} )$, hereafter $\mathcal{L}(\mathbf{\Theta})$, and any priors $p( \mathbf{\Theta} )$.

As befits our data, we employ the log-likelihood for the Poisson point process
\begin{equation}
    \label{eq:loglikelihood-poisson-point-process}
    \log \mathcal{L}(\mathbf{\Theta}) \propto  \sum_{i=0}^{N} \rho(\mathbf{X}_{i}; \mathbf{\Theta}) - \int_{\mathrm{V}} \rho(\mathbf{X}; \mathbf{\Theta} ) \mathrm{d}V
\end{equation}
\noindent where here $\mathbf{X}_{i}$ are coordinate vectors for individual data points. The integral in the likelihood is over the volume $V$ encompassed by the particles.

For each remnant we use all star particles tagged in the last snapshot as a part of the subhalo which contributed to the merger. We place the following domain priors on the fitting of each remnant: $\alpha_{2} > \alpha_{1}$, $\alpha_{2} > 0$, and $\max[\epsilon_{\star},\min(\{ r_{i} \})] < r_{s} < \max(\{ r_{i} \})$. Notably we do not enforce that $\alpha_{1} > 0$ which may seem a natural requirement. While testing we found that some remnants have density profiles which genuinely decrease in density near $r=0$, and others which are very flat near $r=0$ and require a negative value for $\alpha_{1}$ to sufficiently flatten the two-power profile to obtain a good fit. We place logarithmic priors on the values of $\alpha_{2}$ and $r_{s}$ to encourage them not to diverge towards large values (even though $r_{s}$ is bounded).

We draw samples from the posterior using the affine-invariant Markov chain Monte Carlo (MCMC) ensemble sampler of \citet{goodman10} implemented in \texttt{emcee} by \citet{foreman-mackey13}. We initialize our walkers in a small volume of parameter space around a set of parameters determined by optimizing the log-likelihood using the conjugate direction method of \citet{powell64}. We use 100 walkers and draw 2000 samples per walker, discarding 500 per walker as burn-in. We examine by-eye the posterior distributions to ensure that they are well-converged. We do not observe any bimodal or otherwise unduly complicated posterior distributions, and therefore define the best-fitting parameters as the median of the distribution, with the uncertainties being defined as the 16th and 84th percentiles.

Next we determine the anisotropy-specifiying parameters for the DFs, either the single value for the constant-anisotropy DF, $r_{a}$ for the Osipkov-Merritt DF, or $\{ r_{a,1},r_{a,2},k_\mathrm{om} \}$ for the superposition Osipkov-Merritt DF. To obtain these parameters we directly fit the anisotropy, defined according to equation~\eqref{eq:beta}, as a function of radius. We compute the anisotropy by binning all star particles according to the prescription at the beginning of \S~\ref{sec:fitting-distribution-functions} and computing the radial, azimuthal, and polar velocity dispersions. Note that we also test whether using the mean-square velocities, as opposed to the velocity dispersions, affects our fits, and find minimal impact. This is important for remnants with net rotation as the dispersions are not equivalent to the mean-squares.

We define a simple Gaussian objective function $\mathcal{O}$ for binned data to find the best-fitting parameters
\begin{equation}
    \label{eq:beta-objective-function}
    \log \mathcal{O}(\Theta) = \sum_{i=0}^{N} -\frac{ \big[ \beta_\mathrm{M}(r_{i};\Theta) - \beta_{i} \big]^{2} }{2/m_{i}}\,,
\end{equation}
\noindent where $\Theta$ is the parameters which specify the anisotropy, $\beta_\mathrm{M}(r_{i};\Theta)$ is the model anisotropy evaluated at a radial bin center $r_{i}$, $\beta_{i}$ are the remnant velocity dispersions computed as described above, and $m_{i}$ is the total amount of mass in stars per bin (even though the number of particles per bin is constant their mass can vary). 

The model anisotropy is trivial for the constant-anisotropy DF, and for the standard Osipkov-Merritt DF it is given by equation~\eqref{eq:beta-osipkov-merritt}. For the superposition Osipkov-Merritt DF determining the model anisotropy is not so simple, and computation of the individual velocity dispersion profiles for each component DF is necessary. The radial and tangential $(\sigma_{t} = [\sigma_{\phi}^{2} + \sigma_{\theta}^{2}]^{1/2})$  velocity dispersions for the superposition Osipkov-Merritt model are given by \citet{merritt85} as
\begin{equation}
\label{eq:superposition-osipkov-merritt-dispersions}
\begin{split}
    \sigma_{r}^{2} = &\ k_\mathrm{om} \sigma_{r,1}^{2} + (1-k_{\mathrm{om}}) \sigma_{r,2}^{2} \\      
    \sigma_{t}^{2} = &\ k_\mathrm{om} \sigma_{t,1}^{2} + (1-k_{\mathrm{om}}) \sigma_{t,2}^{2}\,, \\      
\end{split}
\end{equation}
where dispersions with subscripts 1 and 2 are computed from the Osipkov-Merritt DFs with $r_{a,1}$ and $r_{a,2}$ respectively.

Again due to the fact that computing DF moments is expensive, it is intractable to obtain these dispersion profiles on-the-fly when fitting. We navigate this issue by computing $\sigma_{r}$ and $\sigma_{t}$ on a grid of $r$ and $r_{a}$ for each remnant. We consider 10 values of $r_{a}$ logarithmically spaced from $10^{-1}$~kpc to $10^{2.5}$~kpc and 20 values of $r$ logarithmically spaced from $\max[\epsilon_{\star},\min(\{ r_{i} \})]$ to $\max(\{ r_{i} \})$. Using interpolation we can then quickly compute velocity dispersions, and then the anisotropy using equation~\eqref{eq:beta}, for any $r_{a}$ at any radius while fitting. We examine the velocity dispersion grids by-eye, and find them to appear well-behaved and smooth, such that we are comfortable using interpolation in this situation.

To fit we follow the same approach as used for the density profile above, first optimizing the objective function to obtain initial conditions and then using MCMC to sample the posterior. Again, we examine our posteriors and find no complicated features or multimodalities and so define the best-fitting anisotropy, $r_{a}$, and $\{ r_{a,1},r_{a,2},k_\mathrm{om} \}$ for each model respectively as the medians of the respective posterior distributions.

Finally we fit for the rotating component of the DF. We determine this in a DF-independent manner by fitting for the asymmetry between positive and negative values of $L_{\mathrm{z}}$. For the prescription outlined in \S~\ref{subsec:rotating-df-prescription} the asymmetry between positive and negative values of $L_{\mathrm{z}}$ is
\begin{equation}
    \label{eq:Lz-anisotropy}
    A(L_{\mathrm{z}}) = \frac{k_\mathrm{rot}}{2} \tanh(L_{\mathrm{z}}/\chi) + \frac{1}{2}\,.
\end{equation}
\noindent The asymmetry in $L_{\mathrm{z}}$ is computed for each remnant by evenly binning from $[-L_{\mathrm{z},80},L_{\mathrm{z},80}]$ where $L_{\mathrm{z},80}$ is the 80th percentile of $\lvert L_{\mathrm{z}} \rvert$ with a number of bins equal to the number of star particles divided by 20 (but no less than 50 bins). The asymmetry is computed as the number count in each bin divided by the sum of the number count in the bin and its complement (at opposite $L_{\mathrm{z}}$). When a bin and its counterpart at opposite $L_{\mathrm{z}}$ contain no star particles they are removed from consideration. 

We use the same objective function, equation~\eqref{eq:beta-objective-function}, as was used for the anisotropy data, however we replace the measured anisotropy in bins and predicted anisotropy with the measured and predicted $L_{\mathrm{z}}$ asymmetry, and change the mass term in the denominator to the mass in each $L_{\mathrm{z}}$ bin. We follow the same approach as was used to determine the best-fitting parameters for the anisotropy profile, again finding no complex features in the posteriors and therefore using the median $k_\mathrm{rot}$ and $\chi$ as the best-fitting parameters.

\begin{figure*}
    \centering
    \includegraphics{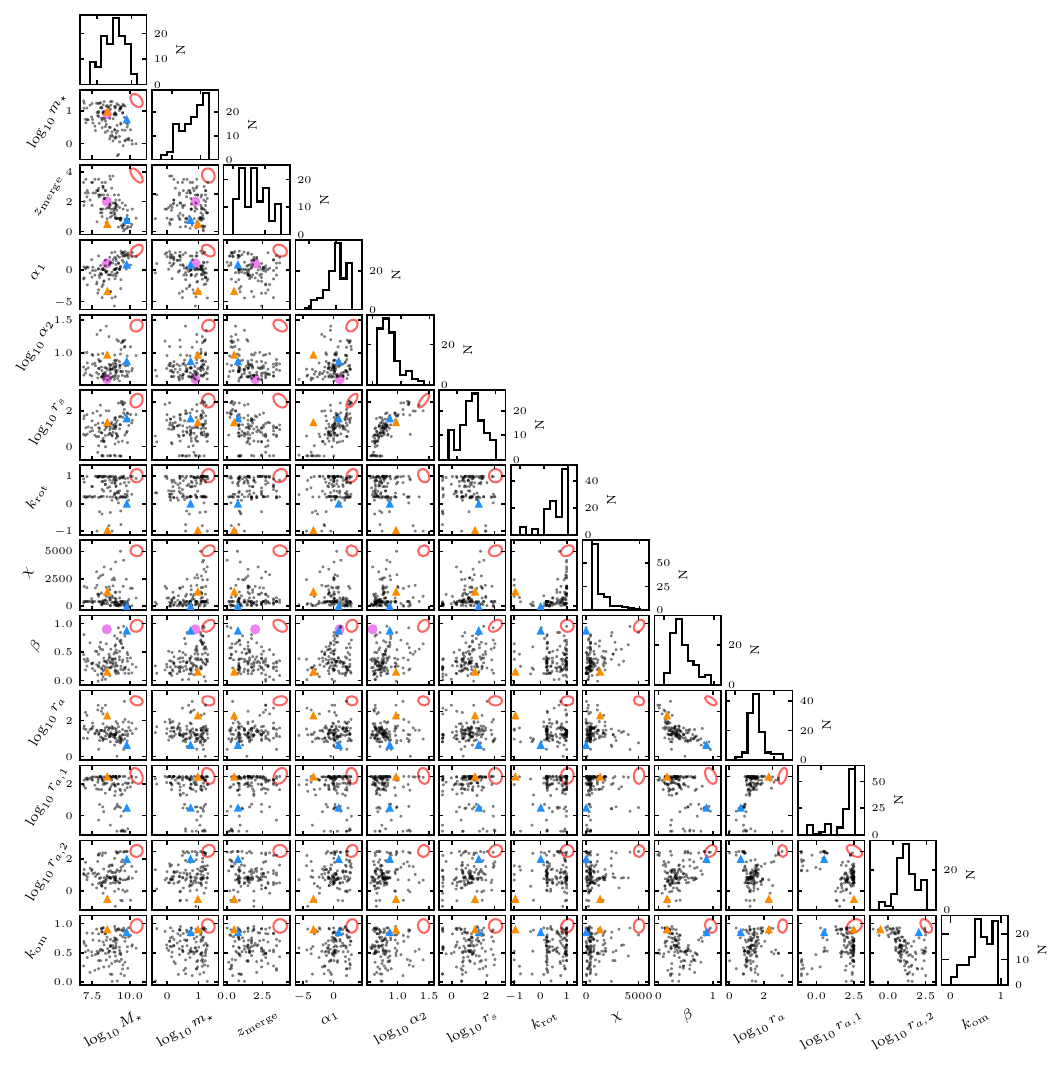}
    \caption{Corner plot summarizing the properties of the 116 major mergers as well as the best-fitting density profile and DF parameters for each of the three classes of DF considered. Shown first are the merger properties: log secondary stellar mass in \Msun, log major-to-minor stellar mass ratio, and merger redshift. Following that are the remnant density profile parameters: inner power law index, log outer power law index, and log of the scale radius in kpc. Then is the rotational fraction and rotation scale parameter. Finally are the DF parameters: anisotropy for the constant-anisotropy DF, log scale radius for the single Osipkov-Merritt model in kpc, log scale radii for the combined Osipkov-Merritt models in kpc, and the mixture fraction for the combined Osipkov-Merritt model. The red ellipse in the corner of each panel is a representation of the covariance matrix for those parameters. The purple circle is the approximate properties of the GS/E remnant (see text for details). The blue and orange triangles are the remnants chosen to be GS/E and Sequoia case studies respectively.}
    \label{fig:density-df-params}
\end{figure*}

With all of the ingredients for our DFs determined individually we build each DF out of its constituent components by solving for $f_{1}(\mathcal{E})$, $f(Q)$, or $f_{1}(Q)$ and $f_{2}(Q)$ pursuant to the definitions in \S~\ref{subsec:distribution-function-models} and then applying the rotation prescription. Figure~\ref{fig:density-df-params} is a corner plot showing the best-fitting density profiles and DF properties derived in this section. Also shown are remnant stellar masses, merger stellar mass ratios, and merger redshifts for reference and comparison. In the top corner of each panel is an ellipse representing the covariance matrix of those parameters, with axis ratios being the ratios of the eigenvalues of the matrix, and the orientation set by the eigenvectors. For context, we show the properties of the GS/E remnant as a large purple circle in the background in relevant panels (i.e. where the properties of the remnant are known). We assume the following for GS/E: the anisotropy to be $\beta=0.9$ \citep{belokurov18,lancaster19}; the inner and outer power laws to be $[1,4]$ with a break radius of 20~kpc \citep[Approximately representing the findings of][]{han22,lane23}; the accretion redshift to be $z=2$ \citep[see][]{mackereth19a,montalban21}; the stellar mass to be $3\times10^{8}$ \citep[a value typical of recent findings;][]{mackereth20,han22,lane23}; and the stellar mass ratio to be 1:8 \citep[As estimated by][]{lane23}. 

A number of noteworthy trends present themselves here beyond the relations among stellar mass, mass ratio, and merger redshift already apparent in Figure~\ref{fig:merger-params}. First we see that remnant stellar mass correlates with inner power law index $\alpha_{1}$ such that more massive remnants tend to have steeper inner density profiles. Remnant anisotropy does not exhibit a strong correlation with remnant stellar mass, however high-anisotropy remnants do tend to have moderate to large stellar masses $\gtrsim 10^{8.5}$~\Msun. But the remnants with the largest stellar masses $\gtrsim 10^{9.5}$~\Msun\ actually tend to have modest anisotropies of $0 < \beta < 0.5$. Remnant anisotropy more strongly correlates with merger stellar mass ratio such that more even mergers $(m_{\star} \sim 1-5)$ tend to me more ergodic than uneven mergers $(m_{\star} \sim 10-20)$. This is an interesting observation given the current prevailing paradigm that the Milky Way was subject to a recent ($z_\mathrm{merger} \sim 2$), significant merger, the remnant of which is the GS/E population in the stellar halo which exhibits a high-degree of radial anisotropy. Remnant stellar mass finally appears to correlate with $r_{a}$, however, because $r_{a}$ and $\beta$ are intimately linked this is similar to noting that stellar mass correlates with anisotropy.

Merger redshift tends to correlate with quantities which also trend with stellar mass such (i.e. density profile parameters and anisotropy), because these two quantities are intimately linked. In a hierarchical Universe all structures, including smaller dwarf galaxies, grow with time and therefore mergers are larger at later times. But in addition to the quantities which also trend with stellar mass, merger redshift appears to trend with remnant net rotation such that early mergers typically (always for $z>2.5$) co-rotate with $k_\mathrm{rot} > 0$, while only late-time mergers exhibit $k_\mathrm{rot} < 0$. This is likely a reflection of the fact that early mergers play a key role in setting the overall angular momentum in a growing galaxy, and therefore early mergers will tend to match the angular momentum of the primary at $z=0$. Of note is the general lack of retrograde mergers, only 9 out of 116 major mergers have $k_\mathrm{rot} < 0$.

Among the density profile parameters $\alpha_{1}$, $\alpha_{2}$, and $r_{s}$ all correlate. Given the correlations with stellar mass noted above, this indicates that more massive mergers tend to have steep inner power laws and strong breaks at large radii, while lower mass mergers tend to have shallow inner power laws with weak breaks and shallow outer power laws. Constant anisotropy $\beta$ is strongly anticorrelated with $r_{a}$, which is understandable given that smaller $r_{a}$ implies that the Osipkov-Merritt DF will have a high anisotropy over a larger radial range. There are also relations between the three superposition Osipkov-Merritt parameters and other parameters, but to note these is mostly to revisit the already noted trends including the other anisotropy parameters.

\section{Assessment and comparison of distribution functions}
\label{sec:assessment-comparison-distribution-functions}

Two tasks now lie before us. The first is to first determine whether these DFs provide a satisfactory model for the remnant N-body data. The second is to gauge whether one DF is superior to the others, and if so under which conditions or for which type of remnants. To tackle the first question we will compare the N-body data to each best-fitting DF model in a self-consistent manner. For the second question we will compare the DF models to each other in the context of the underlying N-body data.

When directly comparing models with N-body data to answer the first question posed above we will proceed in a numerical manner. This is facilitated by drawing samples from each of the best-fitting DFs for each remnant to act as a representation of the DF. The advantage of sampling each DF, as opposed to computing the metrics we will summarize in the next section in a different numerical manner, is that sampling intrinsically accounts for shot noise and allows for error estimation in the metrics through bootstrapping. For each remnant, and each DF fit to that remnant, we draw a number of samples equal to the number of N-body star particles in the remnant from the respective DF. We sample radii between the minimum and maximum N-body particle radii. For more information on how we generate samples from DFs see Appendix~A4 of \citet{lane23}.

\subsection{Two case studies: GS/E and the Sequoia}
\label{subsec:case-studies}

Before moving to a broader comparison of all remnants, which will necessitate generating summary statistics that may cloud the specific details of any individual remnant, we first examine two case studies. We select remnants matched to two of the best-studied stellar halo populations in the Milky Way: GS/E and Sequoia. GS/E is a comparatively massive remnant with high anisotropy and low net rotation \citep{belokurov18,helmi18}, while Sequoia is thought to be a lower mass remnant with high net rotation in the direction opposite disk rotation \citep{myeong19,naidu20}. We record the properties of the selected merger remnant analogs in Table~\ref{tab:merger-case-study-properties}. For more information about merger tree identifiers see the Illustris-TNG documentation\footnote{\url{https://www.tng-project.org/data/docs/specifications}}. The properties of the chosen mergers are highlighted in Figure~\ref{fig:density-df-params}.

\begin{table}
    \centering
    \caption{Key identifying characteristics of the GS/E and Sequoia case study merger remnants. SID is short for SubfindID, and is the identification for a specific subhalo within the TNG50-1 simulation merger tree at a given snapshot (here the $z=0$ snapshot). MLPID is short for MainLeafProgenitorID, which is the unique identification of the branch of the merger tree corresponding to the secondary. The notation $0^N$ indicates $N$ repeating zeros, employed for brevity. $M_{\star}$ is secondary stellar masses and $m_{\star}$ is the primary to secondary mass ratio. Both are reported at the epoch where the secondary reaches its maximum stellar mass. $z_\mathrm{merger}$ is the redshift of the last snapshot where the secondary appears as a distinct subhalo.}
    \begin{tabular}{lcc}
        Property & GS/E & Sequoia \\
        \hline \\
        Primary $z=0$ SID    & 522530 & 518682 \\
        Secondary MLPID      & 9601673 &  $20^{6}280^{3}41936$ \\ 
        $M_{\star}$ [\Msun]  & $6.2\times10^{9}$ & $3.2\times10^{8}$ \\
        $m_{\star}$          & 5.6:1 & 9.6:1 \\
        $z_{\mathrm{merge}}$ & 0.79 & 0.52 \\
    \end{tabular}
    \label{tab:merger-case-study-properties}
\end{table}

To select a GS/E analog we examine remnants with $\beta > 0.8$ (numbering 6 total), and choose one with stellar mass of $\approx 6\times10^{9}$~\Msun\ and anisotropy of $\beta=0.88$, which was deposited in a 6:1 stellar mass ratio merger (30:1 dark matter mass ratio) at $z \approx 0.8$. In choosing the GS/E analog we focus most on anisotropy, attempting to match the value for GS/E of $\sim 0.9$. The stellar mass of GS/E has been subject to a wide range of estimates, spanning nearly two decades from $10^{8}$ to $10^{10}$~\Msun, however more direct measurements of the density profile of GS/E have tended to settle on a lower stellar mass range of $1.5-7.2\times10^{8}$~\Msun\ \citep{mackereth20,han22,lane23}. These estimates typically correspond with a total mass ratio for the merger in the range of 1:4--1:10 (normally assuming a GS/E merger epoch of $z\approx2$). So the mass ratios are in broad agreement with estimates for GS/E, while the total stellar mass is on the high-end of estimates. The accretion epoch for GS/E is thought to be 7-11~Gyr ago ($z \sim 0.8-2.4$), and so this merger is on the very lowest end of that range. This partially explains the higher total stellar mass. Only one of the candidate remnants with high-anisotropy merged at $z>1$, however it only has a present-day stellar mass of only $\sim 10^{7}$~\Msun, much too low to be considered an appropriate analog for GS/E.

For the Sequoia analog we examine strongly counter-rotating remnants with $k < -0.5$ (also numbering 6 total). We select one with a strong degree of rotation, $k=-0.96$, and stellar mass $3\times10^{8}$~\Msun, which is comparable to current estimates \citep{myeong19,matsuno19,naidu20}. The remnant has low anisotropy ($\beta=0.14$), and merged comparatively late ($z=0.52$) compared with the Sequoia remnant which is though to have been deposited between $z=1-2$ \citep{kruijssen20}. The typical angular momentum $L_\mathrm{z}$ for Sequoia ranges between $-3$ and $-1 \times 10^{3}$~kpc~km~s$^{-1}$, and our chosen remnant spans a slightly lower range with 16th, 50th, and 84th percentile $L_\mathrm{z}$ of $-2.3,-1.1$ and $0.2\times 10^{3}$~kpc~km~s$^{-1}$.

\begin{figure*}
    \centering
    \includegraphics[width=\textwidth]{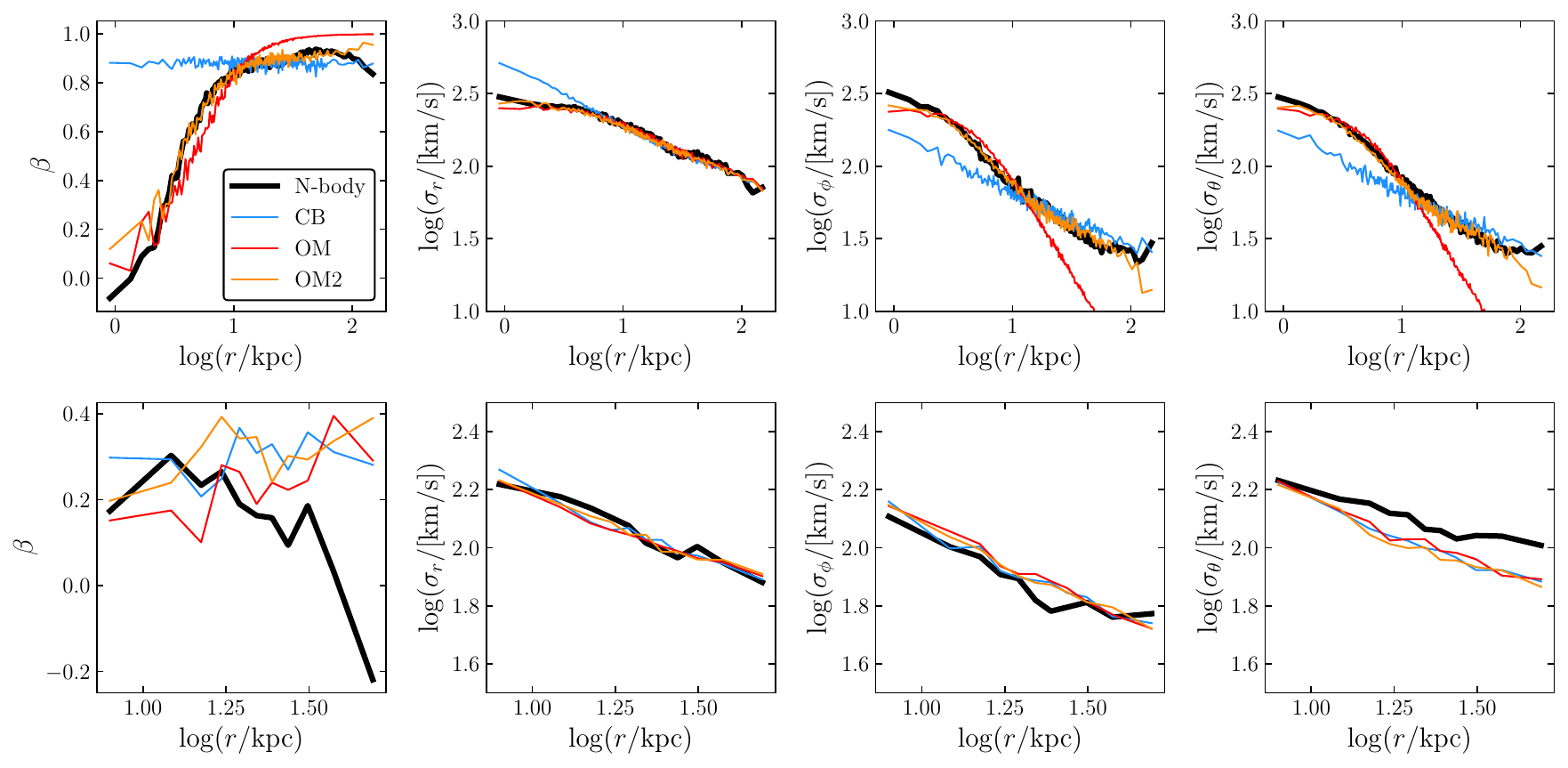}
    \caption{Anisotropy and velocity dispersions for the two case studies from \S~\ref{subsec:case-studies}, the GS/E (top row) and Sequoia (bottom row). The columns, from left to right, show as functions of radius: anisotropy, radial velocity dispersion, azimuthal velocity dispersion, and polar velocity dispersion. In each panel the black curve shows the N-body data, and the blue, red, and orange curves show the samples from the best-fitting constant-anisotropy, standard Osipkov-Merritt, and superposition Osipkov-Merritt DFs respectively.}
    \label{fig:case_study_beta_vdisp}
\end{figure*}

Figure~\ref{fig:case_study_beta_vdisp} shows the velocity dispersions and anisotropy profiles for the GS/E (top row) and Sequoia (bottom row) case studies. For each quantity the N-body data is shown in black, with the DF sample realizations shown in blue (constant-anisotropy), red (Osipkov-Merritt), and orange (superposition Osipkov-Merritt). Here we also point out a minor pathology in our fitting procedure. The rotating DF prescription for strongly rotating remnants (i.e. our Sequoia analog) tends to decrease $\sigma_{\phi}$ in DF samples, because $L_\mathrm{z}$ values are biased towards being similar (at positive or negative values depending on the sense of rotation). While the absolute values of $L_{z}$ remain the same, and therefore overall distribution of orbit shapes is conserved, the concentration of $L_{z}$ to positive or negative values decreases the spread in tangential velocity, which decreases $\sigma_{\phi}$. This effect can be seen in the bottom row of Figure~\ref{fig:case_study_beta_vdisp} where the sample anisotropies seem slightly larger than would fit the data. Strangely, the sample $\sigma_{\theta}$ values are actually lower than the N-body data while the $\sigma_{\phi}$ values are larger than the data, where one might assume the opposite given the effect of the rotation prescription. This is due to an intrinsic mismatch between the $\sigma_{\theta}$ and $\sigma_{\phi}$ values for this remnant specifically, and a keen-eyed reader can indeed see that the sample $\sigma_{\phi}$ values are lower than the sample $\sigma_{\theta}$ values by about 0.1 dex.

To characterize the impact of the rotation perscription on sample anisotropy we fit constant and Osipkov-Merritt anisotropy profiles to the sampled data in the same manner as was done in \S~\ref{subsec:fitting-constant-anisotropy-osipkov-merritt-dfs}. Comparing the constant anisotropy and Osipkov-Merritt scale radius values between the fits to the data and the model samples, we find that indeed the anisotropy does tend to be overestimated. The typical absolute increase in anisotropy is about $\Delta \beta = [0.05,0.1]$ for $\lvert k_{rot} \rvert = [0.5,1.0]$. The Osipkov-Merritt scale radius tends to decrease among rotating samples, with fractional differences being about $\Delta r_{a} = [-5,-25]$~per~cent for $\lvert k_{rot} \rvert = [0.5,1.0]$. The median absolute difference in $r_{a}$ is $1.36$~kpc. These effects could be mitigated by fitting the anisotropy and rotational profile iteratively, or perhaps in a joint manner, but this would be impractical and likely require by-eye assessment of the fits to each remnant. Since the observed biases are minor we simply note this effect and continue with our analysis.

Returning to Figure~\ref{fig:case_study_beta_vdisp}, for the GS/E analog we see that the constant-anisotropy DF correctly specifies the characteristically large anisotropy found at large radii ($10-100$~kpc), but at radii less than 10~kpc the anisotropy profile softens until it approaches zero at the smallest radii. This results in an interesting interplay between the DFs, whereby the Osipkov-Merritt model more correctly models the lower anisotropy at small radii, but overestimates the anisotropy at large radii, because the model always tends towards $\beta=1$. The constant-anisotropy model, on the other hand, more accurately tracks the large anisotropy at large radii, but obviously cannot accommodate the trend towards isotropy at small radii. These differences are mirrored in the individual velocity dispersion trends, which also show each model performing well at either small or large radii. On the other hand this situation exemplifies the benefits of the superposition Osipkov-Merritt model, which is able to reproduce the anisotropy along with each velocity dispersion well at all radii.

For the Sequoia analog, which has a much lower typical anisotropy of about 0.2, all models are able to reasonably capture the behaviour at all radii, keeping in mind the bias due to the rotation prescription described above. For the Osipkov-Merritt DF the best-fitting $r_{a}$ is very large ($\sim 500$~kpc), which allows the anisotropy and velocity dispersion profiles to approach constant values. Note that this is only possible because the analog has a constant, yet low, value. Were the value to be much higher the Osipkov-Merritt model would struggle to replicate its constancy. Notably the tangential velocity dispersions are fit reasonably, but as noted above this is likely due to an intrinsic mismatch in the azimuthal and polar velocity dispersions for this specific remnant, since the rotation prescription should cause the azimuthal dispersions to be underestimated. The radial velocity dispersions are fit well by the model at all radii.

\begin{figure}
    \centering
    \includegraphics[width=\columnwidth]{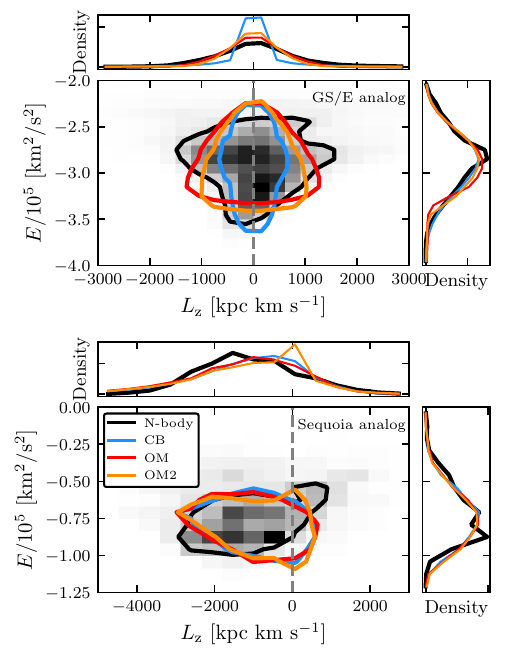}
    \caption{Energy and z-axis angular momentum for the GS/E analog (top panels) and Sequoia analog (bottom panels). The black contours show the N-body data, while the blue, red, and orange contours show the samples from the best-fitting constant-anisotropy, standard Osipkov-Merritt, and superposition Osipkov-Merritt DFs respectively. The background histogram shows the binned N-body data. For each analog the N-body and DF samples contain the same number of data points, are binned identically, and the contours rest at the same levels. The smaller panels above and to the right of each primary panel show the density margins of the angular momentum and energy respectively, binned in the same manner as the primary panels.}
    \label{fig:case_study_ELz}
\end{figure}

Figure~\ref{fig:case_study_ELz} shows energy and $L_\mathrm{z}$ for the N-body data (black histogram), as well as the DF samples from the best-fitting models (histograms not shown). The energies for the N-body data are offset by the median energy of the particles within 5~per~cent of the stellar half-mass radius, calculated in the interpolated potential (the potential which governs the DFs) in order to bring the energies in line with the DF samples. A single contour is shown for each distribution, placed at the same density level for the N-body data as for the DF sample data (recall they have the same number of data points) in order to compare the distributions.

For the GS/E analog the N-body data displays a downward-pointing triangular morphology, centered about $L_\mathrm{Z}=0$, which is typical for both the stellar halo in aggregate, but also for individual accretion remnants. The constant-anisotropy DF spans a very narrow range of angular momenta, which reflects the fact that all orbits must have low $L_\mathrm{z}$ to enforce the high anisotropy at all radii. The Osipkov-Merritt samples display the opposite behaviour compared with the N-body data, where $L_\mathrm{z}$ tends to decrease with increasing energy, resulting in an upward-pointing triangular morphology. This can be understood in the context of the model, where the anisotropy at large radii (roughly equivalent to large energies) is high, therefore requiring angular momentum to be low. Conversely at small radii (roughly equivalent to low energies) the angular momentum spans a larger range, because the anisotropy is low. The superposition Osipkov-Merritt model does not do much better than the standard Osipkov-Merritt, only taking a slightly narrower profile. Overall each model is able to appropriately capture the correct range of energies (about $10^{3}$~km$^{2}$~s$^{-2}$), and the Osipkov-Merritt models are able to appropriately capture the range of angular momentum (roughly $\pm 1500$~kpc~km~s$^{-1}$) unlike the constant-anisotropy model. But none of the models are able to replicate the specific upward-pointing triangular morphology exhibited by the N-body data.

For the Sequoia analog the story is broadly similar. Here, each model is able to correctly capture the locus of the N-body distribution in angular momentum, as well as the spread in energies. Their similarity in this regard is expected given that they share the rotational DF prescription. Along the same terms as Figure~\ref{fig:case_study_beta_vdisp}, each DF model has become similar in nature to fit the constant, low anisotropy, which is reflected in the similarity of their sample distributions. But the actual morphology of the N-body distribution, where energy decreases as angular momentum increases, is not well-fit by the DF samples, which exhibit the opposite trend. This trend exhibited by the DF samples is as-expected for low-anisotropy rotating models, whereby angular momentum is larger for orbits with larger energies. So perhaps this particular Sequoia analog defies these expectations, which could link back to the specifics of the merger.

To summarize this exercise, generally these simple DF models are able to accurately capture the broad strokes of the kinematics of N-body data to which they are fit: spans in energy and angular momentum, net rotation, and the magnitudes of the velocity dispersion profiles. But the specifics of the data are not well-fit, excepting the case of the superposition Osipkov-Merritt model GS/E analog anisotropy and dispersions, which are in good agreement. However, the detailed morphology in the energy-angular momentum plane is not captured well by any of the models.

\subsection{Metrics for DF assessment and comparison}

In this subsection we outline three metrics that we employ both for comparison among the best-fitting DFs, as well as comparison between N-body data and the samples representing the DFs. The first metric is based on the Jeans equation, and is constructed to gauge the suitability of the N-body data for equilibrium modelling. The second metric is a weighted comparison of velocity dispersion profiles, and the third metric uses the likelihood for the Poisson point process. These latter two are designed to determine which DF fits the data better, and to estimate whether the models fit the data well or poorly.

\subsubsection{The Jeans equation}
\label{subsubsec:jeans-equation}

The first metric we consider is based on the time-indepent Jeans equation in spherical coordinates
\begin{equation}
    \label{eq:jeans-equation-spherical}
    \frac{\mathrm{d} (\nu\,\overline{v^2_r})}{\mathrm{d} r} +\,\nu\,
    \left(\frac{\mathrm{d} \Phi}{\mathrm{d} r}+
    \frac{2\overline{v_r^2}-\overline{v_\theta^2}-\overline{v_\phi^2}}{r}\right) = 0\,,
\end{equation}
where $\nu$ is the number density of the tracer, $\overline{v^{2}}_{[r,\phi,\theta]}$ are the spherical mean-square velocities (equivalent to the squared velocity dispersions when the mean velocity in that dimension is zero), $\Phi$ is the underlying gravitational potential, and $r$ is the spherical radius. For steady-state spherical systems in equilibrium this equation is satisfied, and the right-hand-side is equal to zero. Computing this equation for a system in disequilibrium will result in a non-zero residual related to the time derivative of the DF. 

We can compute these terms in the spherical Jeans equation for our N-body data, and then examine the magnitude of the residual to gauge whether the remnant we are interested in appears to be in equilibrium or not. Because computation of the Jeans equation as laid out in equation~\eqref{eq:jeans-equation-spherical} is DF-independent, this does not allow us to make statements about how well each DF fits the data, but rather it may add crucial information about the state of equilibrium to inform our investigation of the merits of other DFs.

We first ``normalize'' the Jeans equation by dividing it by a characteristic term $\nu\overline{v^{2}_{r}}/r$ such that the equation and any residual is unitless. This allows us to compare residuals from different DFs on a more even footing. We hereafter refer to residuals of the Jeans equation normalized in this way as $\mathcal{J}$. We compute the terms of the Jeans equation numerically as follows. We first define a set of bins that will be used to compute the radial derivatives of $\nu \overline{v^{2}_{r}}$ and $\Phi$. These bins are defined as per the beginning of \S~\ref{sec:fitting-distribution-functions}. We then define a second set of bins for the undifferentiated terms $\nu$ and the anisotropy term within the brackets of equation~\eqref{eq:jeans-equation-spherical}. The edges for these bins are the bin centers for the radial derivative term bins. In this way each set of quantities is computed, and then the radial derivatives are evaluated numerically at positions corresponding to the bin centers for the undifferentiated quantities.

\begin{figure}
    \centering
    \includegraphics[width=\columnwidth]{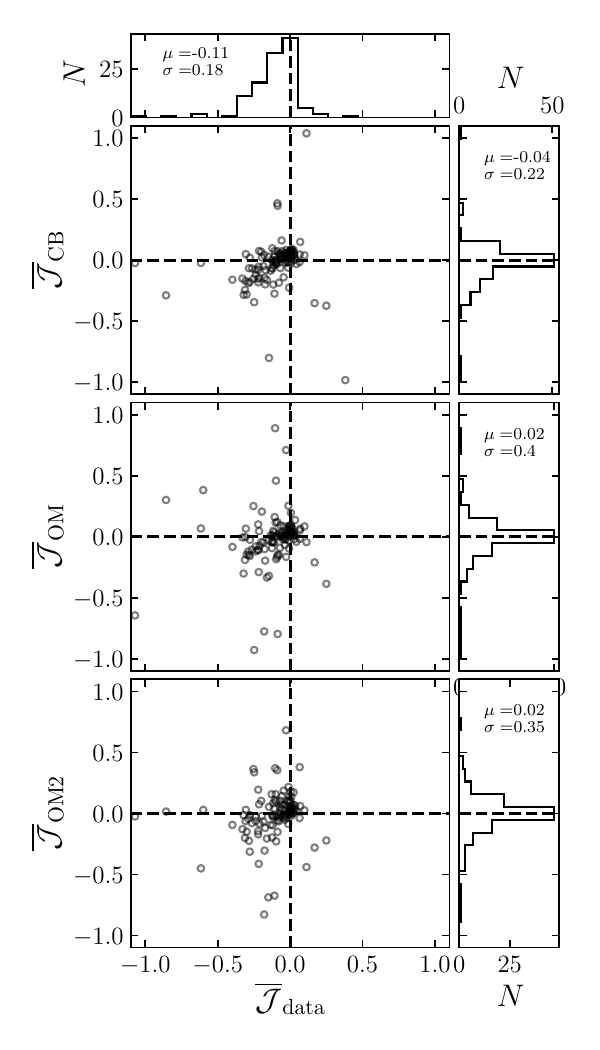}
    \caption{Mass weighted mean $\mathcal{J}$ for each remnant studied in this work. Of the primary panels, the top shows the $\overline{\mathcal{J}}$ for the samples from the best-fitting constant beta DF versus the N-body data. The middle and bottom show the same but for the best-fitting Osipkov-Merritt and superposition Osipkov-Merritt DFs respectively. The dashed lines mark $\overline{\mathcal{J}}=0$. The top-most panel shows the marginal distribution of $\overline{\mathcal{J}}_{\mathrm{data}}$ and the two rightmost panels show the same for each set of DF samples respectively.}
    \label{fig:J-mean}
\end{figure}

\begin{figure}
    \centering
    \includegraphics[width=\columnwidth]{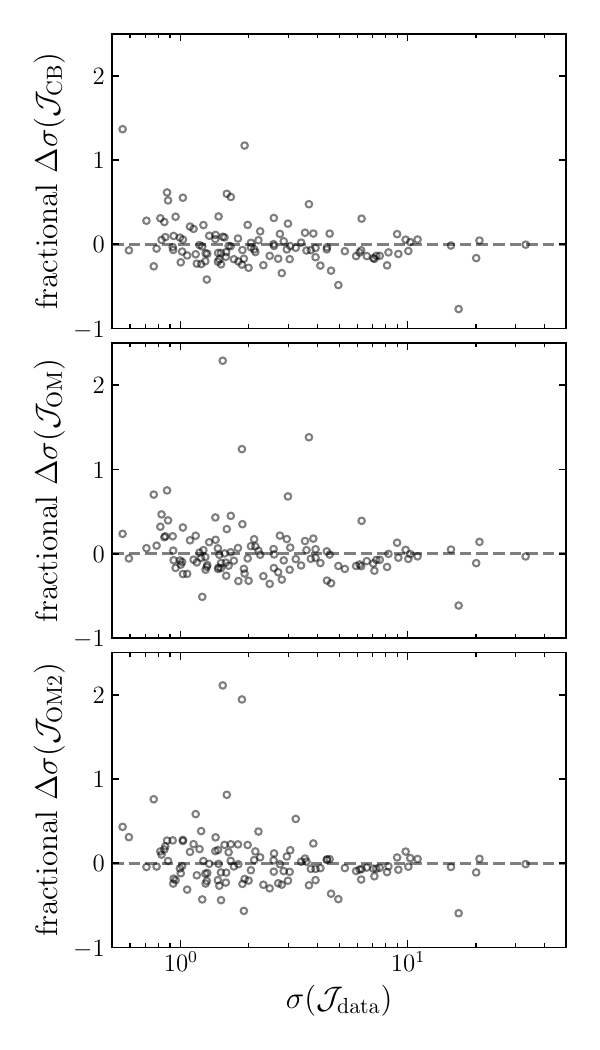}
    \caption{Mass weighted $\mathcal{J}$ root-mean-square about $\overline{\mathcal{J}}$, $\sigma(\mathcal{J})$, for each remnant studied in this work. Each panel shows the fractional difference between the value of $\sigma(\mathcal{J})$ computed with the N-body particles and the corresponding DF sample. The panels are laid out in the same way as Figure~\ref{fig:J-mean}. The dashed line shows equal values of $\sigma(\mathcal{J})$ for N-body particles and DF samples.}
    \label{fig:J-dispersion}
\end{figure}

We are left with the residual of the Jeans equation computed on a radial grid. We compute two metrics to summarize the residuals, the first, $\overline{\mathcal{J}}$ is the mean of the residuals across all bins, weighting by the mass of stars in each bin (note that while each bin contains an equal number of star particles, the masses of the star particles can vary). The second, $\sigma(\mathcal{J})$, is the root-mean-square of $\mathcal{J}$ about $\overline{\mathcal{J}}$, also weighted by the mass of stars in each bin. We compute these quantities both for the stellar N-body data for each remnant, as well as the samples drawn from each best-fitting DF. Comparisons of the quantities $\overline{\mathcal{J}}$ and $\sigma(\mathcal{J})$ between N-body data and DF samples is shown in Figures~\ref{fig:J-mean} and \ref{fig:J-dispersion} respectively.

Because the samples drawn from the best-fitting DFs are in equilibrium and satisfy the spherical Jeans equation by definition, these two figures strongly suggest that the remnants we study are also in equilibrium. First, the values of $\overline{\mathcal{J}}$ for the DF samples shown in Figure~\ref{fig:J-mean} clearly cluster nicely around zero. We would not expect the values to be exactly zero even if the Jeans equation is satisfied, because these are discrete realizations of the DF. The N-body data shows surprisingly comparable behaviour, with the data clustering about zero and exhibiting a scatter comparable to the DF sample data. The N-body data exhibits a slight negative bias, with a mean of $-0.13$, which is larger in magnitude than the means of the DF sample data. But this offset is smaller than the scatter in the distribution, which itself is not larger than the scatter for either set of DF sample data.

Figure~\ref{fig:J-dispersion} shows us that the magnitude of the scatter of the Jeans residual about the weighted mean is comparable for both N-body data and model. One might expect much larger values of $\sigma(\mathcal{J})$ for the N-body data, but this does not appear to be the case. Together, these data strongly suggest that it is reasonable to consider the N-body remnants to be in equilibrium, at least when compared with a comparable discrete realization of an explicitly time-independent DF. But using these results we cannot discern which DF provides a better fit to the data.

\subsubsection{Velocity dispersion profiles}
\label{subsec:velocity-dispersion-profiles}

The second metric we construct compares two sets of velocity dispersion profiles. The velocity dispersions for the N-body data are the same as was computed in \S~\ref{subsec:fitting-constant-anisotropy-osipkov-merritt-dfs}, via binning using the standard method. Velocity dispersion and anisotropy profiles for the model samples are computed similarly, using the same set of radial bins employed for the N-body data dispersions and anisotropies. The model sample velocity dispersions are not computed on their own grid because it is important for the metric we will now describe that the two sets of dispersions are on the same radial grid.

We then compute
\begin{equation}
    \label{eq:velocity-dispersion-delta}
    \delta = \frac{ \sum_{i}^{N} \big[ \lvert \sigma_{\rm data}(r_{i}) - \sigma_{\rm model}(r_{i}) \rvert m_{i} / s_{i} \big] }{ \sum_{i}^{N} m_{i} }\,,
\end{equation}
\noindent where $\sigma_{\rm data}$ are the velocity dispersions of interest ($r$, $\phi$, or $\theta$) for N-body data in radial bins centered on $r_{i}$, and $\sigma_{\rm model}$ are the sampled model velocity dispersions in the same bins. $s_{i}$ is the uncertainty in the N-body data velocity dispersions determined by taking the difference between the 84th and 16th percentiles of a sample of 100 bootstrap realizations of the velocity dispersion computation, with fixed bin sizes. Finally, $m_{i}$ is the mass of N-body star particles in each velocity dispersion bin, which is included because even though each bin contains the same number of particles they can have differing masses. In words this metric expresses the mass-weighted sum of the difference between N-body data and DF model sample velocity dispersions, scaled by the uncertainty in the N-body data.

We compute this metric, comparing N-body data and model samples for each DF, for radial, azimuthal, and polar velocity dispersions. We also compute the metric for the anisotropy profile in the same way, using equation~\eqref{eq:velocity-dispersion-delta} but substituting $\beta$ for $\sigma$. Because $\delta$ relies on binning it is sensitive to the discreet nature of the underlying data and resulting shot noise. We gauge the impact of this by computing $\delta$ for the DF samples against themselves using 100 bootstrap realizations. We record the median and central 68th percentile of the distribution of resulting values of $\delta$. This benchmark is the equivalent of an uncertainty limit on the value of $\delta$ and lends helpful context to the computed values in the sense that it represents the standard for a fit which is indistinguishable from the DF, given the number of data points.

\begin{figure*}
    \centering
    \includegraphics[width=\linewidth]
    {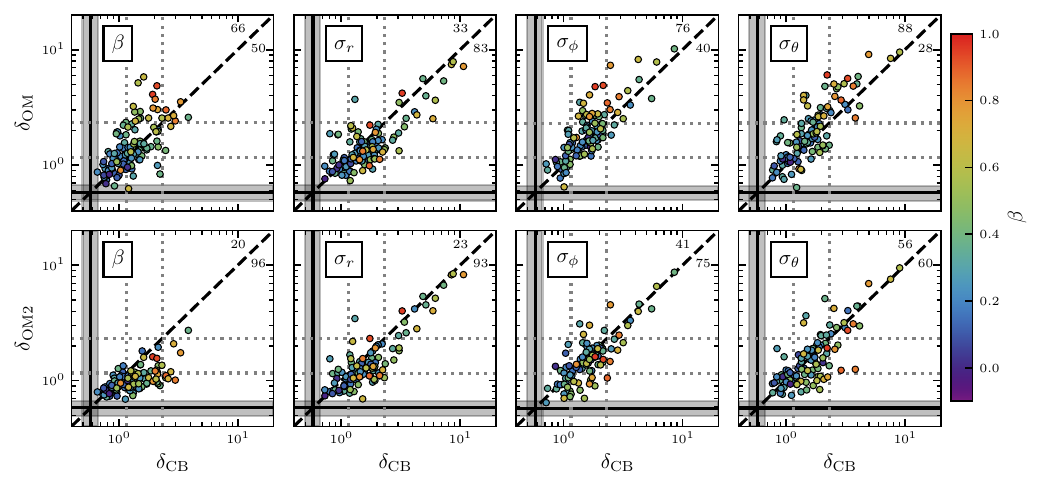}
    \caption{Comparison of $\delta$, the mass- and uncertainty-weighted deviation between data and model, calculated for $\beta$ and $\sigma_{[r,\phi,\theta]}$ (labelled, left to right). $\delta$ is determined using N-body data against the best-fitting DF samples. Delta is computed with the Osipkov-Merritt DF samples (top row) and linear combination Osipkov-Merritt DF samples (bottom row), and is shown against delta computed with the constant-anisotropy DF samples. The colour of the points shows the best-fitting constant anisotropy for the remnant. The solid lines and the surrounding fill show the median and central 68~per~cent interval of the bootstrapped computation of $\delta$ comparing each set of DF samples against itself. The dashed lines are placed at 2 and 4 times the median. The thick dashed line shows the 1:1 relation for reference. The numbers in the top-right corner of each panel show the number of points in the corresponding half of the figure defined by the 1:1 line.}
    \label{fig:delta-comparison}
\end{figure*}

Figure~\ref{fig:delta-comparison} compares the values of $\delta$ determined using the constant-anisotropy DFs and the Osipkov-Merritt DFs. Because larger values of $\delta$ indicate a larger deviation between N-body data and the respective DF samples, comparing $\delta$ in this way reveals whether one DF fits the data better than another. Also shown on Figure~\ref{fig:delta-comparison} are solid lines representing the median value of $\delta$ computed for the DF samples against themselves. Dashed lines are placed at 2 and 4 times this value. 

Examining this figure reveals some interesting trends. First is that $\delta$ computed using anisotropy profiles appears to favour the constant-anisotropy DF compared with the standard Osipkov-Merritt DF. Interestingly, this holds for both high- and low-anisotropy remnants. The velocity dispersion profiles show a different trend. For radial velocity profiles the Osipkov-Merritt profiles are favoured over the constant-anisotropy models. Whereas for the azimuthal and polar velocity dispersions the constant-anisotropy models are again favoured. This can be understood by referring back to Figure~\ref{fig:case_study_beta_vdisp}, which showed the velocity dispersion and anisotropy profiles for the GS/E and Sequoia case studies. There we saw that the radial velocity dispersion is well-fit over all radii by the Osipkov-Merritt model, and in particular it is able to capture the lower dispersions at small radii exhibited by both analog data. The Osipkov-Merritt DF struggles somewhat with the azimuthal and polar velocity dispersions, however, and cannot fully capture the overall anisotropy profile, especially for the highly anisotropic GS/E analog. The high-anisotropy remnant radial velocity dispersions appear to be better fit by the Osipkov-Merritt model, while the tangential velocity dispersions are not, which would track inline with this narrative.

On the other hand, the superposition Osipkov-Merritt DF again demonstrates its effectiveness as it exhibits values of $\delta$ which are much lower than the corresponding value for the constant-anisotropy DF across the board. The difference is most stark for the anisotropy and the radial velocity dispersion, and less for the tangential velocity dispersions. This can again be understood by referring to Figure~\ref{fig:case_study_beta_vdisp}, where we see that the superposition Osipkov-Merritt model is well-matched in all dispersions as well as anisotropy, in contrast to the constant-anisotropy model. These results clearly suggest that the superposition Osipkov-Merritt model should be favoured in comparison with the other two simpler models.

\subsubsection{Model log likelihoods}
\label{subsubsec:model-loglikelihoods}

The final metric we consider is a comparison of the total log likelihood of each model, computed via a modified version of equation~\eqref{eq:loglikelihood-poisson-point-process} where the density $\rho$ is replaced with the value of the corresponding DF, $f$, and the volume integral over configuration space is replaced with a volume integral over the relevant portion of 6D phase space (i.e. energy and angular momentum for the constant-anisotropy and Osipkov-Merritt DFs). We compute the total log likelihood using the N-body data for each DF ($\mathcal{L}_\mathrm{CB}$, $\mathcal{L}_\mathrm{OM}$, and $\mathcal{L}_\mathrm{OM2}$ for the constant-anisotropy, standard Osipkov-Merritt, and superposition Osipkov-Merritt models respectively).

\begin{figure}
    \centering
    \begin{subfigure}{\linewidth}
        \includegraphics[width=\linewidth]{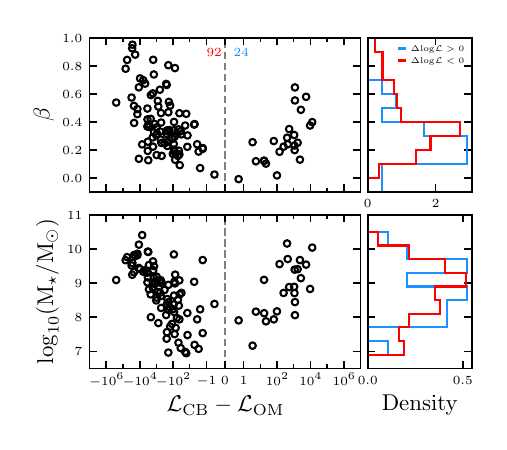}
    \end{subfigure}
    \begin{subfigure}{\linewidth}
        \includegraphics[width=\linewidth]{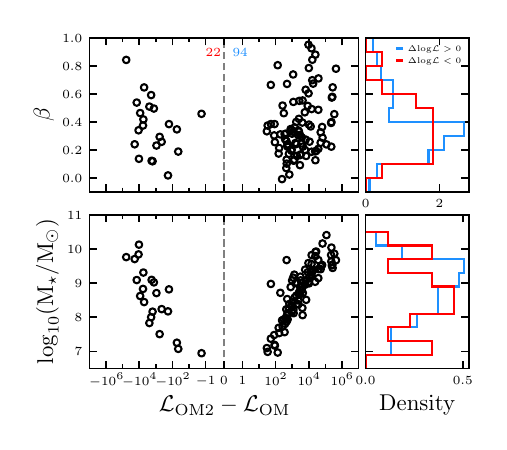}
    \end{subfigure}
    \caption{Difference between the total log likelihood for the constant-anisotropy and Osipkov-Merritt models (top) and linear combination Ospikov-Merritt and standard Osipkov-Merritt (bottom). The axes are log-symmetric outside the range of -1 to 1, which is linear. Of each set of main panels, the top shows the difference in log likelihood as a function of anisotropy, while the bottom shows logarithmic stellar mass of the merger remnants. The right panels show the margins of the two quantities, split into groups where the difference in log likelihoods is greater than 0 and less than 0 (blue and red respectively) and is shown as density. Numbers in each of these samples are shown in blue and red on the main panels.}
    \label{fig:loglike-diff-beta-starmass}
\end{figure}

Figure~\ref{fig:loglike-diff-beta-starmass} shows the difference between the log likelihoods computed for the constant-anisotropy and Osipkov-Merrit DFs against the stellar mass of the remnant as well as the best-fitting constant anisotropy. The figure is replicated below it but for the standard Osipkov-Merritt against the superposition Osipkov-Merritt. The scale of each abscissa is log-symmetric, with the range from -1 to 1 being linear. The axes on the right-hand side show margins for the dependent quantities, split by whether $\Delta \mathcal{L}$ is greater than (blue) or less than (red) zero.

In regard to the top set of panels first, we see the majority of points lie on the left-hand side of the figure (92 versus 24) such that the likelihood for the Osipkov-Merritt model is greater than the constant-anisotropy model. Additionally, we see a trend such that the magnitude of the difference in the log likelihood tends to increase as both remnant mass and anisotropy increase. This is expected because the total log likelihood is a sum over all data points, and therefore the difference would be expected to increase with higher remnant mass. From a probabilistic perspective, more data points means higher confidence in the choice of one model over another. 

When examining the distributions split about $\mathcal{L}_\mathrm{CB} - \mathcal{L}_\mathrm{OM} = 0$ we see that for stellar mass they are broadly similar, with similar peaks and distribution widths. For anisotropy, however, it is clear that the distribution with $\mathcal{L}_\mathrm{CB} - \mathcal{L}_\mathrm{OM} < 0$ is peaked at higher values of anisotropy and has a tail which extends to the highest values of $0.8 < \beta < 1$. This indicates that while, in general, the likelihood analysis favours the Osipkov-Merritt model, it is specifically good at fitting remnants with high anisotropy.

Now moving on to the bottom set of panels, which compare standard Osipkov-Merritt against the superposition version of the DF. We see that here the majority of the points now lie on the right side of the figure, indicating preference for the superposition Osipkov-Merritt model. Again, we can investigate whether there are noteworthy trends with anisotropy by comparing the distributions divided by $\Delta \mathcal{L}$ in terms of anisotropy and stellar mass. We see less evidence here for a preference of the superposition DF among high-anisotropy remnants, although this conclusion is anchored by a single remnant with high-anisotropy which is better fit by the standard Osipkov-Merritt model. However it is clear that the distribution of points with $\Delta \mathcal{L} > 0$ is peaked around anisotropy of 0.4. This is expected given that standard Osipkov-Merritt can struggle to fit these intermediate anisotropy remnants, because its anisotropy prefers to toggle between $\beta = 0$ and $\beta = 1$, whereas the superposition model is able to plateau at intermediate values of $\beta$ over extended radial ranges.

One additional factor that we might consider here is whether or not the preference for the superposition Osipkov-Merritt DF over the standard version is driven by the increased number of parameters in the model (three versus one), which could hint at overfitting. We could consider that if we were to compare the Bayesian information criteria for each model it would be equivalent to adding a factor of $\sim \ln(N) k / 2$ to the difference of likelihoods already shown. Because the number of particles is the same in each instance, ranging roughly between $10^{4}-10^{6}$, and comparing $k=3$ and $k=1$ we have a range of about $9-14$ by which the value of $\Delta \mathcal{L}$ would be modified. Because most values of $\Delta \mathcal{L}$ for which the superposition DF is preferred are greater than $10^{2}$ and none are less than $10$ we can infer that we are not overfitting the problem by adding the two parameters for the superposition DF, and that the superposition DF should be genuinely preferred in most cases.

\section{Discussion}
\label{sec:discussion}

\subsection{The properties of merger remnants around Milky Way analogs}

While not the principal goal of this work, we have isolated and studied 116 merger remnants around 30 Milky Way analogs, fitting density profiles and distribution functions to them. We are able to report a large number of interesting relations, most of which are summarized in Figure~\ref{fig:density-df-params}. First, and certainly not unexpected in the context of $\Lambda$CDM, is the relationship between merger redshift and secondary stellar mass, which is well-defined but too broad (about two decades in secondary stellar mass at fixed redshift and a factor of two in redshift at fixed secondary stellar mass) to be of much use in the context of inference in the Milky Way stellar halo. We also find that the only retrograde remnants $k_\mathrm{rot} < 0$ merged at redshifts $z<2.5$, and they are few in number (9 out of 116). We attribute this to the fact that early mergers are probably key in setting the angular momentum of the host galaxy, and therefore we would expect $k_\mathrm{rot} > 0$. But interestingly there is an overall lack of retrograde remnants even at redshifts $z<2.5$.

In regard to density profile parameters, we find that secondary stellar mass (and therefore merger stellar mass ratio, because the Milky Way analog sample has a narrow range of stellar masses) correlates well with the inner slope of the remnant density profile: more massive remnants are steeper and less massive remnants shallower. Additionally the density profile parameters are well-correlated, such that remnants with steeper inner power laws have sharper transitions at larger radii when compared with remnants with shallow inner power laws which tend to also have shallow outer power laws. Connecting these trends with the relation between merger stellar mass and density profile parameters paints a broader picture. Lower mass remnants tend to have more uniform, shallow density profiles, while higher mass remnants have steeper density profiles with sharp cutoffs. These results are in line with the findings of \citet{deason18}, who posit that the sharp density cutoff observed in the Milky Way stellar halo at $r \sim 20$~kpc is due to apocenter pileup of debris from a single large progenitor (in this case GS/E).

\subsection{A comparison of the DF models}

\subsubsection{The constant anisotropy and Osipkov-Merritt DFs}

Our findings with respect to the three DFs studied in this work are mixed. First, we clearly see that the superposition Osipkov-Merritt DF model exhibits superior performance among the metrics by which we gauge DF performance as well as the more detailed investigations of the GS/E and Sequoia analogs. It is able to more accurately reproduce the velocity dispersion and anisotropy profiles of a wider range of remnants than either the constant-anisotropy or standard Osipkov-Merritt DF models. In addition the likelihood analysis clearly favours this model over the other two, even when taking into consideration the fact that it requires additional parameters.

But frustratingly none of the DF models were able to fully reproduce the observed morphology of either the GS/E or Sequoia analog remnants in the energy-angular momentum space. While each model was able to do a reasonable job of getting the locus and extent in energy and angular momentum of the N-body particles, they were unable to capture some of the semantic details in the distributions, such as the specific trends in the width and location of the distribution of angular momentum with varying energy. These failures highlight that while these models can be effective they are clearly limited. Coupling this with the fact that the DFs --- particularly the superposition Osipkov-Merritt DF --- can match the dispersions and anisotropy of the remnants well suggests that higher order moments of the DF could play an important role.

With this in mind it does appear to be the case that high-anisotropy remnants are better represented using the Osipkov-Merritt (and therefore naturally the superposition Osipkov-Merritt as well) model than the constant-anisotropy DF. The constant-anisotropy model substantially overpredicts the velocity anisotropy via the radial velocity dispersion at small radii. This leads to an overprediction of the density of stars with low angular momenta, as is seen in Figure~\ref{fig:case_study_ELz}.

\begin{figure*}
    \centering
    \includegraphics{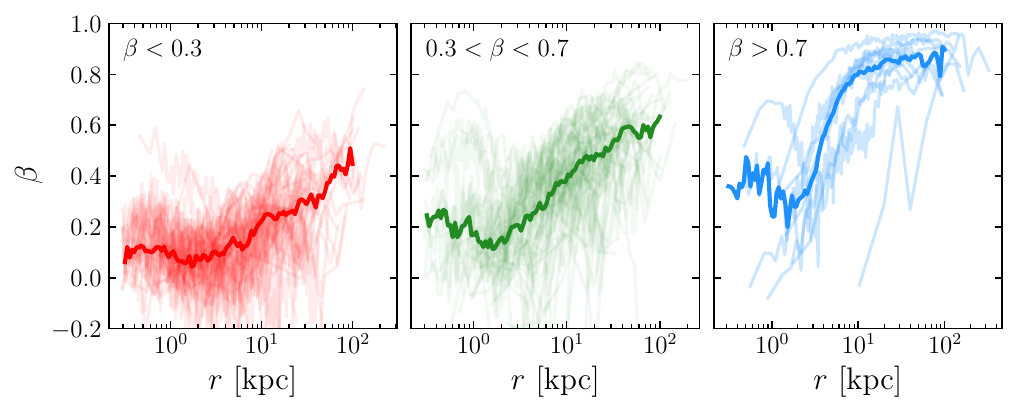}
    \caption{A comparison of the anisotropy profiles of the remnants studied in this work. Each panel shows the radial anisotropy profile for remnants based on their best-fitting constant anisotropy, from left to right: $\beta < 0.3$, $0.3 < \beta < 0.7$, and $\beta > 0.7$. See \S~\ref{subsec:fitting-constant-anisotropy-osipkov-merritt-dfs} for details on anisotropy profile computation. Superimposed on each set of profiles is a trend computed by treating each individual profile as a linear interpolation (necessary because the anisotropy for each remnant is computed using a unique set of bins) and calculating the median at a range of radial positions from each interpolator at that point.}
    \label{fig:anisotropy-profile-comparison}
\end{figure*}

Building on this conclusion that Osipkov-Merritt-like DFs provide a better fit to high anisotropy remnants, we ask the following question: do the anisotropy profiles of such remnants actually resemble the Osipkov-Merritt profile or is it simply the best among the available DFs? To answer this question, we show in Figure~\ref{fig:anisotropy-profile-comparison} the individual anisotropy profiles for each remnant, divided as a function of the best-fitting constant anisotropy. First, we see that the remnants with the highest average anisotropy clearly have Osipkov-Merritt-like anisotropy profiles, which are low in the inner galaxy and plateau at large radii. The evolution of anisotropy profile shape in each of the panels of Figure~\ref{fig:anisotropy-profile-comparison} is also interesting. All anisotropy profiles appear to rise at some point, and the remnants with lower average anisotropy have shallower transitions that reach lower values of $\beta$. This trend is also reflected in Figure~\ref{fig:density-df-params}, the distribution of best-fitting DF parameters, where we can see a strong anticorrelation between best-fitting $r_{a}$ and constant $\beta$. Figure~\ref{fig:anisotropy-profile-comparison} gives context to that relationship: remnants with lower typical anisotropy which plateaus at lower values require large $r_{a}$ to accommodate the gentle rise in anisotropy (i.e. $\beta$ only approaches 1 far outside the radial range of the remnant). Remnants with high anisotropy on the other hand have more dramatic increases in anisotropy that more directly resemble the Osipkov-Merritt profile, and require smaller $r_{a}$ so that the anisotropy profile can rise sharply.

\subsubsection{Consideration for other DFs}
\label{subsubsec:consideration-for-other-dfs}

Given the perceived shortcomings of the models considered in this work, highlighted by the analysis of the GS/E and Sequoia analogs in \S~\ref{subsec:case-studies}, it is natural to consider other options. First and most significantly are action-based DFs such as the models put forward by \citet{binney14d} and built upon by \citet{posti15}. In these models the DF is a function of the actions $[J_{r},J_{z},L_{z}]$. They support a wide range of density profiles, and naturally incorporate variations in anisotropy, rotation, and flattening. One downside of these models is that the density and anisotropy profiles are not directly specified. Furthermore, they require computation of the actions, which is typically done using an approximation such as `St\"{a}ckel fudge' \citep{binney12,mackereth18b} that can exhibit pathologies for highly radial orbits or those which move far from the disk-plane \citep[see][]{mackereth18b}. This contrasts with the energy and angular momentum which are always well-defined. Action-based models have been recently used by \citet{gherghinescu24} to model the stellar halos (not individual merger remnants) of M31 analogs in the Auriga simulations, for which they found such models are well-suited. They have also been used to model the whole Milky Way stellar halo \citep{li22}, the distribution of halo globular clusters \citep[][]{callingham22}, as well as dark matter halos \citep{hattori21}. It will be useful in the future to consider a comparison between these action-based DFs and the Osipkov-Merritt DFs for the purposes of modelling individual accretion remnants.

It may also be useful to further explore linear combinations of Osipkov-Merritt DFs as outlined by \citet{merritt85}, which to the best of our knowledge have not been widely applied to the modelling of stellar halos. These DFs can be constructed such that weights for various $r_{a}$ are defined by distributions, offering much more flexibility than the two-$r_{a}$ model we consider. Additionally, DFs with different $r_{a}$ may trace different density profiles, while we only consider components with the same density profile. Such models, however, may be ultimately limited when compared with action-based DFs in the same way as a standard Osipkov-Merritt DF: they are only functions of the energy and angular momentum.

Another promising avenue for future study are neural network driven methods for obtaining DFs such as \textit{Deep Potential} described by \citet{green23}. This method employs normalizing flows, which are able to generate complex distributions via a series of well-defined transformations acting on a simple starting distribution, such as the normal distribution. \citet{green23} demonstrated that such models are able to recover known analytic DFs based on sampled phase space distributions. Another option is to use diffusion-based approaches, where the transformation from a target distribution to a sample distribution (e.g. the normal distribution) via the gradual addition of noise (diffusion) is learned and then inverted. These techniques may be ideally suited for application to the complicated and often messy phase space distributions we observe in the remnants in this work. Additionally, these types of models are likely easier to compute on-the-fly, an issue that we had to work around with our DFs by building them up piece-by-piece.

In this study we focus on fitting the density profile, underlying potential, and anisotropy profile of each remnant to create our DFs. A different approach could be to focus exclusively on energy and angular momentum (in addition possible to other integrals of motion) and attempt to find functional forms that describe their distributions. These would be valid DFs, being functions of integrals of motion, but would not likely correspond to density profiles and potentials of the functional forms we are familiar with (e.g. the two-power models we use in this study). But these DFs would be useful in that they would, by design, be well matched to integrals of motion distributions, which are the common spaces used to study the Milky Way stellar halo at present. This type of strategy may be best suited as a tailored approach to studying specific remnants such as GS/E in the Milky Way stellar halo, since it may not be possible to find functional forms that can be broadly applied. Yet, it is not clear that the DFs we use in this study are of broad use either, and so working solely in integrals of motion space may ultimately prove just as effective if not better.

Finally, in this work we neglect triaxiality both in the underlying potential and the density profile for each remnant. The potential of the Milky Way is certainly flattened in the inner parts of the Galaxy thanks to the stellar disk, and in the outer parts of the Galaxy the potential-governing dark matter halo may be oblate and mildly triaxial \citep{law10,veraciro13,bovy16}. The physical geometry of most remnants in the stellar halo has yet to be thoroughly studied and it is likely that most are at least somewhat flattened, following the general shape of the Milky Way potential, and some may be triaxial. The only remnant to have had its 3D shape thoroughly assessed is GS/E, which is found to be modestly triaxial \citep{iorio21,han22,lane23}. Certainly in the broader cosmological context it is expected that halos with a baryonic component are triaxial \citep{dubinski94,abadi10}. Is it therefore important to consider triaxiality among remnants and the host potential in future studies that seek to use DFs. 

The action-based DFs noted above can incorporate flattening by treating the total angular momentum $L$ differently from the vertical component $L_\mathrm{z}$. Additionally, \citet{sanders15c} demonstrated the construction of action-based, triaxial density profiles with self-consistent DFs based on those of \citet{williams15}. But the problem of the general applicability of triaxial DFs is threefold. The first is the models add a great deal of complexity to the problem at hand, both in computational load and the possible necessity for fine tuning. Second, in so far as the models are based on actions they are not always grounded in physical observables, and the ability of the models to match key observables may be limited. Third, the models can be limited by the accessibility of integrals of motion (or their lack of existence). Integrals of motion for a triaxial system can be acquired using approximate \citep{sanders15a} or perturbative techniques \citep{binney18}, or in a more tailored, accurate manner for systems such as the Galactic bar \citep{binney20}, but such approaches are not applicable broadly. The impact of our choice to work only with spherical systems is challenging to gauge, but unlikely to be substantial. Certainly the DFs will not as faithfully reproduce the underlying data as if they did include a triaxial perscription, however, as demonstrated above many observables are well described, in a broad sense by our spherical DFs. It could be the case that some of the discrepencies in the energy and angular momenta examined in \S~\ref{subsec:case-studies} could be caused by the triaxiality of the underlying potential, the density profile of the remnant, or both. Nonetheless we believe that future work should be done to study the applicability of triaxial DFs for the modelling of remnants. 

\subsubsection{The DF for GS/E}

Of great interest in the \textit{Gaia} era is to achieve detailed descriptions of the kinematics and dynamics of merger remnants in the Milky Way stellar halo so as to link them with their progenitors and the merger events which deposited them in the Galaxy. Density modelling has been successfully applied to this endeavour, both in regard to the stellar halo as a whole \citep[e.g.][]{deason19,mackereth20} as well as individual remnants, particularly GS/E \citep[e.g.][]{han22,lane23}. Our findings that the Osipkov-Merritt DFs are superior to the constant-anisotropy DF for modelling high-anisotropy remnants suggests that these DFs should be used to model GS/E in the future. Indeed, a promising future study would be to determine whether the radial velocity dispersion and the anisotropy drops for GS/E stars near the center of the Galaxy, which would be consistent with our findings here for most high-anisotropy remnants. \citet{lancaster19} studied the anisotropy of the stellar halo using Blue Horizontal Branch stars. They are unable to probe inwards of 10~kpc, but the anisotropy of the GS/E-like population in their study does appear to begin to drop slightly at these radii. \citet{iorio21} use a much more numerous sample of RR Lyrae variables, and do indeed find that the anisotropy of the GS/E-like component does drop substantially within 5~kpc. The fraction of the halo composed of the GS/E-like component does also appear to drop in their data, however this may be a circular argument as if the true GS/E anisotropy does drop in the inner galaxy then the fraction would also drop if the GS/E component in these studies is defined as having high anisotropy. But disentangling potentially low-anisotropy GS/E stars in the inner galaxy from the old \textit{in-situ} stellar halo \citep[e.g.][]{belokurov22,rix22} will be important if this is the case. Abundances may be crucial here, as the \textit{in-situ} halo should have different abundances, especially [Al/Fe] \citep[see][]{belokurov22}, when compared with GS/E. Indeed, given that the chemical sequence of GS/E is now well-constrained with high-purity selections \citep[e.g.][]{lane23} it should be possible to study its anisotropy profile using abundance-selected samples. If we assume that the anisotropy drops from 0.9 around $r=5$~kpc to about 0 in the center of galaxy then an Osipkov-Merritt profile with $2 < r_{a} < 4$~kpc should provide an adequate model for GS/E. 

Building on the discussion from the previous section, while we are confident that Osipkov-Merritt DFs are more suitable for anisotropic remnants like GS/E compared with constant anisotropy models, we do not claim to have generated a definitive DF for the remnant. It is certainly important to consider that GS/E has been demonstrated to exhibit triaxiality \citep{han22,lane23}, indicating that a more accurate DF may be one of the action-based forms including a triaxial perscription. Furthermore, it is an open question whether or not GS/E is sufficiently phase-mixed to be considered in equilibrium. On one hand the majority of the remnant occupies the inner Milky Way stellar halo ($< 20~\mathrm{kpc}$) where the dynamical times are $\lesssim 1~\mathrm{Gyr}$, much shorter than the supposed $8-10$~Gyr age of the remnant \citep{montalban21}. On the other hand the initial evidence of the triaxiality of the remnant from \citet{simion19} and \citet{iorio19}---that it appears linked with the Hercules-Aquila cloud and Virgo Overdensity---may also hint that it is not in a relaxed state if these are non-equilibrium structures instead of simply being concentrations along one triaxial axis. Finally, the evidence presented by \citet{chandra23} of GS/E-affiliated substructure in the outter stellar halo ($> 60$~kpc) clearly indicates that the debris is not in equilibrium at those radii. Indeed, a good candidate DF for GS/E, in combination with stellar survey data with a well-understood selection function, could possibly be used to assess whether GS/E is in equilibrium in the inner Milky Way. Regardless, further work is certainly needed to create a definitive DF for GS/E which includes triaxiality and can accurately match other observables such as the anisotropy, velocity dispersions, and density profile across a large range of radii.

\section{Summary and conclusions}
\label{sec:summary-conclusions}

In this work we study how well commonly used classes of DFs are able to describe merger remnants around Milky Way analogs in the IllustrisTNG simulations. We identify 116 major mergers, defined as having a mass ratio greater than 1:20, around 30 Milky Way analogs in the highest resolution TNG50 simulation run. We fit two-power spherical density profiles to remnants in addition to three classes of DF: the constant-anisotropy, standard Osipkov-Merritt, and superposition Osipkov-Merritt DFs.

We compare the DFs with N-body data by sampling from the DFs in a manner consistent with the N-body data and then generating a variety of summary statistics. We use the Jeans equation to gauge whether or not the remnants are in a state of dynamical equilibrium sufficient for modelling using DFs, and we find that they are. We compute the velocity dispersion and anisotropy profiles, and compare them for the N-body data and DF samples. We also compute the likelihoods for each model and compare them. Additionally, we examine two case studies corresponding to remnants matched to the well-studied GS/E and Sequoia populations in the Milky Way stellar halo.

Our main findings are summarized as follows:

\begin{enumerate}
    \item Remnants with higher degrees of anisotropy are more likely to exhibit sharp breaks in their density profiles, and the steepness of the break correlates with the anisotropy. Remnant anisotropy also correlates with remnant stellar mass. This reinforces observational evidence that apocenter pileup of a single large progenitor is responsible for the sharp break in the density of the Milky Way stellar halo.
    \item The remnants that we study appear, to a good approximation, to be in equilibrium, at least when benchmarked against representative samples from an equilibrium DF.
    \item The DF models we consider are able to capture the correct locus and spread in energies and angular momenta for remnants with a variety of properties. Additionally, the magnitudes of the velocity dispersion profiles are well-matched to N-body data. The specific morphology of the remnants in energy-angular momentum space, as well as their velocity dispersion and anisotropy profiles, was not always well-matched by the best-fitting DFs.
    \item We find clear evidence that highly anisotropy stellar populations with $\beta > 0.8$, such as GS/E, are better modelled using the standard Osipkov-Merritt DF as opposed to the constant-anisotropy DF; an even better model is the superposition of two Osipkov-Merritt DFs. While these remnants may have high average anisotropy, the profiles nearly always become more ergodic in the center of the Galaxy. We estimate that an Osipkov-Merritt profile with a scale radius of 2--4~kpc would provide a reasonable DF model for GS/E.
\end{enumerate}

In general our results were inconclusive with regards to providing a clear set of criterion to decide whether to use the constant-anisotropy or Osipkov-Merritt DF. We do see good evidence that high-anisotropy remnants are better-modelled using the Osipkov-Merritt DF. These remnants typically have radial velocity dispersion and anisotropy profiles which soften at small radii to lower anisotropy, behaviour which is better captured by the Osipkov-Merritt DFs compared with the constant-anisotropy DFs. We therefore leave this as a principal conclusion, and otherwise recommend that with regards to other remnants that one proceed on a case-by-case basis. We do find that superposition Osipkov-Merritt DFs are able to replicate a wide variety of anisotropy profiles, and so may be the most generally applicable. In conclusion, we hope that this study provides useful insight for a future in which much of the stellar halo has been explored and the task turns to detailed modelling.

\section*{Acknowledgements}

We thank the referee for their comments, which have certainly improved the quality of the manuscript. JMML and JB acknowledge financial support from NSERC (funding reference number RGPIN-2020-04712). We are very greatful to Ted Mackereth, who helped to conceptualize this project.

\section*{Data Availability}

The IllustrisTNG simulations used in this article are publicly available at \url{https://www.tng-project.org/data/}.



\bibliographystyle{mnras}
\bibliography{manuscript} 



\appendix



\bsp	
\label{lastpage}
\end{document}